
\input harvmac
\newcount\figno
\figno=0
\def\fig#1#2#3{
\par\begingroup\parindent=0pt\leftskip=1cm\rightskip=1cm\parindent=0pt
\baselineskip=11pt
\global\advance\figno by 1
\midinsert
\epsfxsize=#3
\centerline{\epsfbox{#2}}
\vskip 12pt
{\bf Fig. \the\figno:} #1\par
\endinsert\endgroup\par
}
\def\figlabel#1{\xdef#1{\the\figno}}
\def\encadremath#1{\vbox{\hrule\hbox{\vrule\kern8pt\vbox{\kern8pt
\hbox{$\displaystyle #1$}\kern8pt}
\kern8pt\vrule}\hrule}}

\overfullrule=0pt

%
\def\tilde{\widetilde}
\def\bar{\overline}

\font\zfont = cmss10 

\def\bigone{\hbox{1\kern -.23em {\rm l}}}
\def\ZZ{\hbox{\zfont Z\kern-.4emZ}}

\Title{hepth-9503124, IASSNS-HEP-95-18}
{\vbox{\centerline{STRING THEORY DYNAMICS}
\smallskip
\centerline{ IN VARIOUS DIMENSIONS}}}
\smallskip
\centerline{Edward Witten}
\smallskip
\centerline{\it School of Natural Sciences, Institute for Advanced Study}
\centerline{\it Olden Lane, Princeton, NJ 08540, USA}\bigskip
\baselineskip 18pt

\medskip

\noindent

The strong coupling dynamics of string theories in dimension $d\geq 4$ are
studied.  It is argued, among other things, that eleven-dimensional
supergravity arises as a low energy limit of the ten-dimensional
Type IIA superstring, and that a recently conjectured duality
between the heterotic string and Type IIA superstrings controls
the strong coupling dynamics of the heterotic string in five, six,
and seven
dimensions and implies $S$ duality for both heterotic and Type II strings.
\Date{March, 1995}
\newsec{Introduction}

Understanding in what terms string theories should really be formulated
is one of the basic needs and goals
in the subject. Knowing some of the phenomena
that can occur for strong coupling -- if one can know them without
already knowing the good formulation! -- may be a clue in this direction.
Indeed, $S$-duality between weak and strong coupling for the heterotic
string in four dimensions (for instance, see \nref\sen{A. Sen, ``Strong-Weak
Coupling Duality In Four-Dimensional String Theory,'' Int. J. Mod. Phys.
{\bf A9} (1994) 3707, hepth/9402002.}
\nref\schwarz{J. H. Schwarz, ``Evidence For Non-Perturbative String
Symmetries,'' hep-th/9411178.}
\refs{\sen,\schwarz})
really ought to be a clue for a new formulation of string theory.

At present there is very strong evidence for $S$-duality in supersymmetric
field theories, but the evidence
for $S$-duality in string theory is much less extensive.  One motivation
for the present work was to improve this situation.

Another motivation was to try to relate four-dimensional $S$-duality
to statements or phemonena in more than four dimensions.  At first
sight, this looks well-nigh implausible since $S$-duality between electric
and magnetic charge seems to be very special to four dimensions.  So we are
bound to learn something if we succeed.

Whether or not a version of $S$-duality plays a role, one would like
to determine the strong coupling behavior of string theories above
four dimensions, just as $S$-duality -- and its conjectured Type II
analog, which has been called $U$-duality \ref\hull{C. M. Hull and P. K.
Townsend, ``Unity Of Superstring Dualities,''
QMW-94-30, R/94/33.} -- determines the strong coupling limit after toroidal
compactification to four dimensions.\foot{By ``strong coupling
limit'' I mean the limit as the string coupling constant goes to infinity
keeping fixed (in the sigma model sense) the parameters of the
compactification.  Compactifications that are not explicitly
described or clear from the context
will be toroidal.}  One is curious about the phenomena
that may arise, and in addition if there is any non-perturbative
inconsistency in the higher dimensional string theories (perhaps ultimately
leading to an explanation of why we live in four dimensions) it might
show up naturally in thinking about the strong coupling behavior.

\nref\vafa{C. Vafa, unpublished.}
In fact, in this paper, we will analyze the strong coupling
limit of certain string theories in certain dimensions.  Many of
the phenomena are indeed novel, and many of them are indeed related
to  dualities.  For instance, we will argue in section two
that the strong coupling limit
of Type IIA supergravity in ten dimensions is eleven-dimensional
supergravity!    In a sense, this statement
gives a rationale for ``why'' eleven-dimensional supergravity exists,
much as the interpretation of supergravity theories in $d\leq 10$ as low
energy limits of string theories
explains ``why''  these remarkable theories exist.
How eleven-dimensional supergravity fits into the scheme of things
has been a puzzle
since the theory was first predicted
\ref\nahm{W. Nahm, ``Supersymmetries And Their Representations,''
Nucl. Phys. {\bf B135} (1978) 149.}
and constructed \ref\julia{
E. Cremmer, B. Julia, and J. Scherk, ``Supergravity Theory In  11
Dimensions,'' Phys. Lett. {\bf 76B} (1978) 409.}.

Upon toroidal compactification, one can study the strong coupling
behavior of the Type II theory in $d<10$ using $U$ duality, as we will
do in section three.  One can obtain a fairly complete picture, with
eleven-dimensional supergravity as the only ``surprise.''

Likewise, we will argue in section four that the strong coupling
limit of five-dimensional heterotic string theory is Type IIB in six
dimensions, while the strong coupling
limit of six-dimensional heterotic string theory is Type IIA in six
dimensions (in each case with four dimensions as a K3), and the
strong coupling limit in seven dimensions involves
eleven-dimensional supergravity.
These results are based on a relation between
the heterotic string and the Type IIA superstring
in six dimensions that  has been proposed before \refs{\hull,\vafa}. The
novelty
in the present paper is  to show, for instance, that vexing puzzles
about the strong coupling behavior of the heterotic string in five
dimensions disappear if one assumes the conjectured relation of
the heterotic string to Type IIA in six dimensions.
Also we will see -- using a mechanism proposed previously in
a more abstract setting \ref\duff{M. Duff, ``Strong/Weak Coupling Duality
{}From The Dual String'' hep-th/9501030.} -- that the ``string-string duality''
between heterotic and Type IIA strings in six dimensions implies
$S$-duality in four dimensions, so the usual evidence for $S$-duality
can be cited as evidence for string-string duality.

There remains the question of determining the strong coupling dynamics
of the heterotic string above seven dimensions.  In this context,
there is a curious speculation\foot{This idea was considered
many years ago by M. B. Green, the present author, and probably others,
but not in print as far as I know.} that the heterotic string in ten
dimensions with $SO(32) $ gauge group might have for its
strong coupling limit the $SO(32)$ Type I theory.  In section five, we show
that this relation, if valid, straightforwardly determines the
strong coupling behavior of the heterotic string in nine and eight dimensions
as well as ten, conjecturally completing the description of strong
coupling dynamics except for $E_8\times E_8$ in ten dimensions.

\nref\narain{K. Narain,
``New Heterotic String Theories In Uncompactified Dimensions $<10$,''
Phys. Lett. {\bf 169B} (1986) 41; K. Narain, M. Samadi, and E. Witten,
``A Note On The Toroidal Compactification Of Heterotic String Theory,''
Nucl. Phys. {\bf B279} (1987) 369.}
\nref\ginsparg{P. Ginsparg, ``On Toroidal Compactification Of Heterotic
Superstrings,'' Phys. Rev. {\bf D35} (1987) 648.}
\nref\dine{M. Dine, P. Huet, and N. Seiberg, ``Large And Small Radius
In String Theory,'' Nucl. Phys. {\bf B322} (1989) 301.}
\nref\leigh{J. Dai, R. G. Leigh, and J. Polchinski,
``New Connections Between String Theories,'' Mod. Phys. Lett.
{\bf A4} (1989) 2073.}
The possible relations between different theories discussed in this
paper should be taken together with other, better established
relations between different string theories.  It follows from
$T$ duality that below ten dimensions
the $E_8\times E_8$ heterotic
string is equivalent to the $SO(32)$ heterotic string
\refs{\narain,\ginsparg}, and Type IIA is equivalent to Type IIB \refs{\dine,
\leigh}.
Combining these statements with the much shakier relations discussed
in the present paper, one would have a web of connections between the five
string theories and eleven-dimensional supergravity.

After this paper was written and circulated, I learned of
a paper \ref\townsend{P. Townsend, ``The Eleven-Dimensional Supermembrane
Revisited,'' hepth-9501068.} that has some overlap with the contents
of section two of this paper.

\newsec{Type II Superstrings In Ten Dimensions}

\subsec{Type IIB In Ten Dimensions}

In this section, we will study the strong coupling
dynamics of Type II superstrings in ten dimensions.
We start with the easy case, Type IIB.
A natural conjecture has already
been made by Hull and Townsend \hull.  Type IIB supergravity in ten
dimensions has an $SL(2,{\bf R})$ symmetry; the conjecture is that
an $SL(2,{\bf Z})$ subgroup of this is an exact symmetry of the string
theory.\foot{For earlier work on the possible role of the non-compact
supergravity symmetries in string and membrane theory,
see \ref\duffo{M. F. Duff, ``Duality Rotations In String Theory,''
Nucl. Phys. {\bf B335} (1990) 610; M. F. Duff and J. X. Lu,
``Duality Rotations in Membrane Theory,'' Nucl. Phys. {\bf B347} (1990)
394.}.}
  This then would relate the strong and weak coupling limits
just as $S$-duality relates the strong and weak coupling limits of
the heterotic string in four dimensions.

This $SL(2,{\bf Z})$ symmetry in ten dimensions, if valid,
has powerful implications below ten dimensions.
The reason is that in $d<10$ dimensions, the Type II theory
(Type IIA and Type IIB are equivalent below ten dimensions)
is known to have a $T$-duality symmetry $SO(10-d,10-d;{\bf Z})$.
This $T$-duality group does not commute with the $SL(2,{\bf Z})$ that is
already present in ten dimensions, and together
they generate the discrete subgroup of the supergravity
symmetry group that has been called $U$-duality.
\foot{For instance, in five dimensions, $T$-duality is $SO(5,5)$
and $U$-duality is $E_6$.  A proper subgroup of $E_6$ that contains
$SO(5,5)$ would have to be $SO(5,5)$ itself or $SO(5,5)\times {\bf R}^*$
(${\bf R}^*$ is the non-compact form of $U(1)$),
so when one tries to adjoin to $SO(5,5)$ the $SL(2)$ that was already
present in ten dimensions
(and contains two generators that map NS-NS states to RR states and
so are not in $SO(5,5)$)
one automatically generates all of $E_6$.}
Thus, $U$-duality is true in every dimension below ten if the $SL(2,{\bf Z})$
of the Type IIB theory holds in ten dimensions.

\nref\gseiberg{N. Seiberg, ``Electric-Magnetic Duality In Supersymmetric
Non-Abelian Gauge Theories,'' hepth/9411149.}
In the next section we will see that $U$-duality controls Type II
dynamics below ten dimensions.  As $SL(2,{\bf  Z})$
also controls Type IIB dynamics
in ten dimensions, this fundamental duality between
strong and weak coupling controls all Type II dynamics in all
dimensions except for the odd case of Type IIA in ten dimensions.
But that case will not prove to be a purely isolated exception: the basic
phenomenon that we will find in Type IIA in ten dimensions is highly
relevant to Type II dynamics below ten dimensions, as we will see
in section three.  In a way ten-dimensional Type IIA proves
to exhibit the essential new phenomenon in the simplest context.

To compare to $N=1$ supersymmetric
dynamics in four dimensions \gseiberg, ten-dimensional Type IIA is somewhat
analogous to supersymmetric QCD with $3N_c/2>N_f>N_c+1$,
whose dynamics is controlled by an effective infrared theory that does
not make sense at all length scales.  The other
cases are analogous to the same theory with $3N_c>N_f>3N_c/2$, whose dynamics
is controlled by an exact equivalence of theories -- conformal fixed points --
that make sense at all length scales.

\subsec{Ramond-Ramond Charges In Ten-Dimensional Type IIA}

It is a familiar story to string theorists that the string coupling
constant is really the expectation of a field -- the dilaton field $\phi$.
Thus, it can be scaled out of the low energy effective action by shifting
the value of the dilaton.

After scaling other fields properly,
this idea can be implemented in closed string theories by writing
the effective action as $e^{-2\phi}$ times a function that is invariant
under $\phi\to\phi+{\rm constant}$.  There is, however, an important
subtlety here that affects the Type IIA and Type IIB (and Type I) theories.
These theories have massless antisymmetric tensor fields that originate
in the Ramond-Ramond (RR) sector.  If $A_p $ is such a $p$-form field,
the natural gauge invariance is $\delta A_p=d\lambda_{p-1}$,
with $\lambda_{p-1}$ a $p-1$-form -- and no dilaton in the
transformation laws.  If one scales $A_p$ by a power of $e^\phi$,
the gauge transformation law becomes more complicated and less natural.

Let us, then, consider the Type IIA theory with the fields normalized
in a way that makes the gauge invariance natural.  The massless bosonic
fields from the ${\rm (Neveu-Schwarz)}^2$ or NS-NS sector are the dilaton, the
metric tensor $g_{mn}$, and the antisymmetric tensor
$B_{mn}$.  From the RR sector, one has a one-form $A$
and a three form $A_3$.  We will write the field strengths
as $H=dB$, $F=dA$, and $F_4=dA_3$; one alse needs $F_4'=dA_3+A\wedge H$.
The bosonic part of
the low energy effective action can be written $I=I_{NS}+I_{R}$
where $I_{NS}$ is the part containing NS-NS fields only and $I_R$ is
bilinear in RR fields.  One has (in units with $\alpha'=1$)
\eqn\bospar{I_{NS} = {1\over 2}
\int d^{10}x\sqrt g e^{-2\phi}\left(R+4(\nabla\phi)^2
-{1\over 12}H^2\right) }
and
\eqn\ospar{I_R=-\int d^{10}x \sqrt g\left({1\over 2\cdot 2!}F^2+
{1\over 2\cdot 4!}F_4'{}^2\right)
          -{1\over 4}\int F_4\wedge F_4\wedge B.}
With this way of writing the Lagrangian, the gauge transformation laws
of $A$, $B$, and $A_3$ all have the standard, dilaton-independent
form $\delta X=d\Lambda$, but it is not true that the classical
Lagrangian scales with the dilaton like an overall factor of $e^{-2\phi}$.

Our interest will focus on the presence of the abelian gauge field $A$
in the Type IIA theory.  The charge $W$ of this gauge field has the following
significance.  The Type IIA theory has two supersymmetries  in ten
dimensions, one of each chirality; call them $Q_\alpha$ and $Q'_{\dot\alpha}$.
The space-time momentum $P$ appears in the anticommutators $\{Q,Q\}\sim
\{Q',Q'\}\sim P$.  In the anticommutator of $Q$ with $Q'$ it is possible
to have a Lorentz-invariant central charge
\eqn\nuho{\{Q_\alpha,Q'_{\dot\alpha}\}\sim \delta_{\alpha\dot\alpha}W.}
To see that such a term does arise, it is enough to consider the interpretation
of the Type IIA theory as the low energy limit of eleven-dimensional
supergravity, compactified on ${\bf R}^{10}\times {\bf S}^1$.
{}From
that point of view, the gauge field $A$ arises from the components
$g_{m, 11}$ of the eleven-dimensional metric tensor, $W$ is simply
the eleventh component of the momentum, and \nuho\ is part of the
eleven-dimensional supersymmetry algebra.
\foot{The relation of the supersymmetry algebra to eleven dimensions
leads to the fact that both for the lowest level and even for the first
excited level of the Type IIA theory, the states can be arranged in
eleven-dimensional Lorentz multiplets \ref\bars{I. Bars, ``First Massive
Level And Anomalies In The Supermembrane,'' Nucl. Phys. {\bf B308} (1988)
462.}.  If
this would persist at higher levels, it might be related to the
idea that will be developed below.  It would also be interesting
to look for possible eleven-dimensional traces in the superspace
formulation \ref\gates{J. L. Carr, S. J. Gates, Jr., and R. N. Oerter,
``$D=10,\,\, N=2a$ Supergravity in Superspace,'' Phys. Lett. {\bf 189B}
(1987) 68.}.}

In the usual fashion \ref\wito{E. Witten and D. Olive, ``Supersymmetry
Algebras That Include Central Charges,'' Phys. Lett. {\bf B78} (1978)
97.},
the central charge  \nuho\ leads to an inequality between the mass $M$
of a particle and the value of $W$:
\eqn\uho{M\geq c_0|W|,}
with $c_0$ a ``constant,'' that is a function only
of the string coupling constant
$\lambda=e^\phi$, and independent of which particle is considered.
The precise constant with which $W$ appears in \nuho\
or \uho\ can be worked out using the low energy supergravity
(there is no need to worry about stringy corrections as the discussion
is controlled by the leading terms in the low energy effective action,
and these are uniquely determined by supersymmetry).  We will work this
out at the end of this section by a simple scaling argument starting
with eleven-dimensional supergravity.  For now, suffice it to say
that the $\lambda$ dependence of the inequality is actually
\eqn\guho{M\geq {c_1\over\lambda}|W|}
with $c_1$ an absolute constant.  States for which the inequality
is saturated -- we will call them BPS-saturated states by analogy
with certain magnetic monopoles in four dimensions -- are in ``small''
supermultiplets with $2^8$ states, while generic supermultiplets
have $2^{16}$ states.

In the elementary string spectrum, $W$ is identically zero.
Indeed, as $A$ originates in the RR sector, $W$ would have had to be
a rather exotic charge mapping NS-NS to RR states.
However, there is no problem
in finding classical black hole solutions carrying the $W$ charge
(or any other gauge charge, in any dimension).
It was proposed by Hull and Townsend
\hull\ that quantum particles carrying RR charges
arise by quantization of such black holes.
Recall that, in any dimension, charged black holes obey an inequality
$GM^2\geq {\rm const}\cdot W^2$ ($G, M$, and $W$ are Newton's constant
and the black hole mass and charge); with $G\sim \lambda^2$,
this inequality has the same structure
as \guho.  These two  inequalities actually correspond in the sense that an
extreme black hole, with the minimum mass for given charge, is invariant
under some supersymmetry \ref\hug{G. W. Gibbons
and C. M. Hull, ``A Bogomol'ny Bound For
General Relativity And Solitons in $N=2$ Supergravity,''
Phys. Lett. {\bf 109B} (1982) 190.}
and so should correspond upon quantization
to a ``small'' supermultiplet saturating the inequality \guho.

To proceed, then, I will assume that there are in the theory BPS-saturated
particles with $W\not=0$.  This assumption can be justified as follows.
Hull and Townsend actually showed that upon
toroidally compactifying to less than
ten dimensions, the  assumption follows from $U$-duality.  In toroidal
compactification, the radii of the circles upon which one compactifies
can be arbitrarily big.  That being so, it is implausible to have
BPS-saturated states of $W\not=0$ below ten dimensions unless they exist
in ten dimensions; that is, if the smallest mass of a $W$-bearing state
in ten dimensions were strictly bigger
than $c|W|/\lambda$, then this would remain true after compactification
on a sufficiently big torus.

If the ten-dimensional theory has
BPS-saturated states of $W\not= 0$, then what values of $W$ occur?
A continuum of values of $W$ would seem pathological.  A discrete spectrum
is more reasonable.  If so, the quantum of $W$ must be independent
of the string coupling ``constant'' $\lambda$.  The reason is that
$\lambda$ is not really a ``constant'' but the expectation value of
the dilaton field $\phi$.  If the quantum of $W$
were to depend on the value of $\phi$, then the value of the electric
charge $W$ of a particle would change in a process in which $\phi$
changes (that is, a process in which $\phi$ changes in a large region
of space containing the given particle); this would violate conservation of
$W$.

The argument just stated involves a hidden assumption that will now
be made explicit.  The canonical action for a Maxwell field is
\eqn\canonicalaction{{1\over 4e^2}\int d^nx\sqrt g F^2.}
Comparing to \ospar, we see that in the case under discussion
the effective value of $e$ is independent
of $\phi$, and this is why the charge of a hypothetical charged particle
is independent of $\phi$.  If the action were
\eqn\anonicalaction{{1\over 4}\int d^nx\sqrt g e^{\gamma\phi}F^2}
for some non-zero $\gamma$, then the current density would equal
(from the equations of motion of $A$) $J_m=\partial^n(e^{\gamma\phi}F_{mn})$.
In a process in which $\phi$ changes in a large region of space containing
a charge, there could be a current inflow proportional to $\nabla\phi\cdot F$,
and the charge would in fact change.  Thus, it is really the
$\phi$-independence
of the kinetic energy of the RR fields that leads to the statement
that the values of $W$ must be independent of the string coupling constant
and that the masses of charged fields scale as $\lambda^{-1}$.

\nref\strathdee{R. Kallosh, A. Linde, T. Ortin, A. Peet, and A. Van Proeyen,
``Supersymmetry As A Cosmic Censor,'' Phys. Rev. {\bf D46} (1992) 5278.}
Since the classical extreme black hole solution has arbitrary charge $W$
(which can be scaled out of the solution in an elementary fashion),
one would expect, if BPS-saturated
charged particles do arise from quantization of extreme black
holes, that they should possess
every allowed charge.   Thus, we expect BPS-saturated
extreme black holes of mass
\eqn\momss{M={c|n|\over\lambda},}
where $n$ is an arbitrary integer, and, because of the unknown value
of the quantum of electric charge, $c$ may differ from $c_1$ in \guho.

Apart from anything else that follows, the existence of particles
with masses of order $1/\lambda$, as opposed to the more usual
$1/\lambda^2$ for solitons, is important in itself.  It almost
certainly means that the string perturbation expansion -- which is
an expansion in powers of $\lambda^2$ -- will have non-perturbative
corrections of order $\exp(-1/\lambda)$,
 in contrast to the more usual
$\exp(-1/\lambda^2)$ \foot{If there are particles
of mass $1/\lambda$, then loops of those particles should give effects
of order $e^{-1/\lambda}$, while loops of conventional solitons, with
masses $1/\lambda^2$, would be of order $\exp(-1/\lambda^2)$.}.
The occurrence of such terms has been guessed
by analogy with matrix models \ref\shenker{S. Shenker,
``The Strength Of Non-Perturbative Effects In String Theory,''
in the Proceedings of the Cargese Workshop On Random Surfaces, Quantum
Gravity, And Strings (1990).}.

The fact that the masses of RR charges diverge as $\lambda\to 0$
-- though only as $1/\lambda$ -- is important for self-consistency.
It means that these states disappear from the spectrum as $\lambda\to 0$,
which is why one does not see them as elementary string states.

\subsec{Consequences For Dynamics}

Now we will explore the consequences for dynamics of the existence
of these charged particles.

The mass formula \momss\ shows that, when the string theory is weakly coupled,
the RR charges are very heavy.  But if we are bold enough to follow
the formula into strong coupling, then for $\lambda\to\infty$,
these particles go to zero mass.  This may seem daring, but
the familiar argument based on the ``smallness'' of the multiplets would
appear to show that the formula \momss\ is exact and therefore can be
used even for strong coupling.  In four dimensions, extrapolation
of analogous mass formulas to strong coupling has been extremely successful,
starting with the original work of Montonen
and Olive that led to the idea of $S$-duality.
(In four-dimensional $N=2$ theories, such mass formulas
generally fail to be exact \ref\sw{N. Seiberg and E. Witten,
``Electric-Magnetic Duality, Monopole Condensation, And Confinement
In $N=2$ Supersymmetric Yang-Mills Theory,'' Nucl. Phys.
{\bf B426} (1994) 19.}
because of quantum corrections to the low energy effective action.
For $N=4$ in four dimensions, or for Type IIA supergravity in ten
dimensions, the relevant, leading
terms in the low energy action are uniquely determined by
supersymmetry.)

So for strong coupling, we imagine a world in which
there are supermultiplets of mass $M=c|n|/\lambda$ for every
$\lambda$.
These multiplets necessarily contain particles of spin at least
two, as every supermultiplet in Type IIA supergravity in ten dimensions
has such states.  (Multiplets that do not saturate the mass inequality
contain states of spin $\geq 4$.)  Rotation-invariance of the
classical extreme black hole solution suggests\foot{Were the classical
solution not rotationally invariant, then upon quantizing it one would
obtain a band of states of states of varying angular momentum.  One
would then not expect to saturate the mass inequality of an extreme
black hole without taking into account the angular momentum.}
(as does $U$-duality) that
the BPS-saturated multiplets are indeed in this multiplet of minimum spin.

Thus, for $\lambda\to\infty$ we have light, charged fields of spin two.
(That is, they are charged with respect to the ten-dimensional gauge field
$A$.)  Moreover,
there are infinitely many of these.
This certainly does not correspond to a local field theory in ten dimensions.
What kind of theory will reproduce this spectrum of low-lying states?
One is tempted to think of a string theory or Kaluza-Klein theory
that has an infinite tower of excitations.  The only other option, really,
is to assume that the strong coupling limit is a sort of theory that
we do not know about at all at present.

One can contemplate the possibility that the strong coupling limit
is some sort of a string theory with the dual string scale being
of order $1/\lambda$, so that the charged multiplets under discussion
are some of the elementary string states.  There are two reasons that this
approach does not seem promising: (i) there is no known string theory
with the right properties (one needs
Type IIA supersymmetry in ten dimensions, with
charged string states coupling to the abelian gauge field in the
gravitational multiplet); (ii) we do not have evidence for a
stringy exponential proliferation
of light states as $\lambda\to\infty$, but only for a single
supermultiplet for each integer $n$, with mass $\sim |n|$.

Though meager compared
to a string spectrum, the spectrum we want to reproduce
is just about right for a Kaluza-Klein theory.
Suppose that in the region of large $\lambda$, one should think of the
theory not as a theory on ${\bf R}^{10}$ but as a theory on ${\bf R}^{10}
\times {\bf S}^1$.  Such a theory will have a ``charge'' coming
from the rotations of ${\bf S}^1$.  Suppose  that the radius
$r(\lambda)$ of the ${\bf S}^1$ scales as $1/\lambda$ (provided that
distances are measured using the ``string'' metric that appears in
\bospar\ -- one could always make a Weyl rescaling).  Then for large
$\lambda$,
each massless field in the eleven-dimensional theory
will give, in ten dimensions,
for each integer $n$ a single field of charge $n$ and mass
$\sim |n|\lambda$.  This is precisely the sort of spectrum that we want.

So we need an eleven-dimensional field theory whose fields are in one-to-one
correspondence with the fields of the Type IIA theory in ten dimensions.
Happily, there is one: eleven-dimensional supergravity!
So we are led to the strange idea that eleven-dimensional supergravity
may govern the strong coupling behavior of the Type IIA superstring
in ten dimensions.

Let us discuss a little more precisely how this would work.
The dimensional reduction of eleven-dimensional supergravity to
ten dimensions including the massive states has been discussed in some detail
(for example, see \ref\huq{M. Huq and M. A. Namazie, ``Kaluza-Klein
Supergravity In Ten Dimensions,'' Class. Quantum Grav. {\bf 2}  (1985) 293.}).
Here we will be very schematic, just to touch on the points that are
most essential.  The bosonic fields in eleven-dimensional supergravity
are the metric $G_{MN}$ and a three-form $A_3$.
The bosonic part of the action is
\eqn\pilko{I={1\over 2}\int d^{11}x\sqrt G \left(R+|dA_3|^2\right)
          +\int A_3\wedge dA_3\wedge dA_3.}
Now we reduce to 10 dimensions, taking the eleventh dimensions to
be a circle of radius $e^\gamma$. That is, we take the eleven-dimensional
metric to be
$ds^2=G^{10}_{mn}dx^m\,dx^n+e^{2\gamma}(dx^{11}-A_mdx^m)^2$
to describe a ten-dimensional metric $G^{10}$ along with a vector
$A$ and scalar $\gamma$; meanwhile $A_3$
reduces to a three-form which we still call
$A_3$, and a two-form $B$ (the part of the original
$A_3$ with one index equal to 11).  Just for the massless
fields, the bosonic part of the action becomes
roughly
\eqn\upilko{I={1\over 2}\int d^{10}x \sqrt {G^{10}}
\left(e^{\gamma}\left(R+|\nabla\gamma|^2 +|dA_3|^2\right)
 +e^{3\gamma}|dA|^2 +e^{-\gamma}|dB|^2\right)+\dots.}
This formula, like others below,  is very rough and is
only intended to exhibit the powers of $e^{\gamma}$.
The point in its derivation is that, for example, the
part of $A_3$ that does not have an
index equal to ``11'' has a kinetic energy proportional to $e^\gamma$,
while the part with such an index has a kinetic energy proportional to
$e^{-\gamma}$.

The powers of $e^\gamma$ in \upilko\ do not, at first sight,
appear to agree with those in \bospar.  To bring them in agreement,
we make a Weyl rescaling by writing $G^{10}=e^{-\gamma}g$.  Then
in terms of the new ten-dimensional metric $g$, we have
\eqn\piulko{I={1\over 2}\int d^{10}x\sqrt g\left(e^{-3\gamma}
\left(R+|\nabla\gamma|^2+|dB|^2\right)+  |dA|^2+|dA_3|^2+\dots\right).}
We see that \piulko\ now does agree with
\bospar\ if
\eqn\normo{e^{-2\phi}=e^{-3\gamma}.}
In the original eleven-dimensional metric, the
radius of the circle is $r(\lambda)=e^\gamma$, but now, relating
$\gamma$ to the dilaton string coupling constant via \normo, we can write
\eqn\ormo{r(\lambda)=e^{{2\phi\over 3}}= \lambda^{2/3}.}
The masses of Kaluza-Klein
modes of the eleven-dimensional theory
are of order $1/r(\lambda)$ when measured in the metric $G^{10}$,
but in the metric $g$ they are of order
\eqn\kormo{{e^{-\gamma/2}\over r(\lambda)}\sim
\lambda^{-1}.}
Manipulations similar to what we have just seen will be made many
times in this paper.

Here are the salient points:

(1) The radius of the circle {\it grows} by the formula \ormo\
as $\lambda\to\infty$.
This is important for self-consistency; it means that when $\lambda$
is large the eleven-dimensional
theory is weakly coupled at its compactification scale.  Otherwise
the discussion in terms of eleven-dimensional field theory
would not make sense, and we would not know how to improve on it.
As it is, our proposal
reduces the strongly coupled Type IIA superstring to a field theory
that is weakly coupled at the scale of the low-lying excitations,
so we get an effective determination of the strong coupling behavior.

(2) The mass of a particle of charge $n$, measured in the string metric
$g$ in the effective ten-dimensional world, is of order
$|n|/\lambda$ from \kormo.  This is the dependence on $\lambda$
claimed in \guho,
which we have now in essence derived: the dependence
of the central charge on $\phi$ is uniquely determined by the low
energy supersymmetry, so by deriving this dependence in a Type
IIA supergravity
theory that comes by Kaluza-Klein reduction from eleven dimensions,
we have derived it in general.

So far, the case for relating the strong coupling limit of Type IIA
superstrings to eleven-dimensional supergravity consists of the fact
that this enables us to make sense of the otherwise puzzling dynamics
of the BPS-saturated states and that point (1) above worked out correctly,
which was not obvious {\it a priori}.  The case will hopefully get
much stronger in the next section when we extend the analysis
to work below ten dimensions and incorporate $U$-duality, and in section
four when we look at the heterotic string in seven dimensions.
In fact, the most startling aspect of relating strong coupling
string dynamics to eleven-dimensional supergravity is the
Lorentz invariance that this implies between the eleventh dimension
and the original ten.  Both in section three and in section four, we will
see remnants of this underlying Lorentz invariance.

\newsec{Type II Dynamics Below Ten Dimensions}

\subsec{$U$-Duality And Dynamics}

In this section, we consider Type II superstrings toroidally compactified
to $d<10$ dimensions, with the aim of understanding the strong
coupling dynamics, that is the behavior
when some parameters, possibly including the string coupling constant,
are taken to extreme values.

The strong coupling behaviors of Type IIA and Type IIB
seem to be completely different in ten dimensions, as we have seen.
Upon toroidal compactification below ten dimensions,
the two theories are equivalent under $T$-duality \refs{\dine,\leigh}, and so
can be considered together.  We will call the low energy supergravity
theory arising from this compactification Type II supergravity in $d$
dimensions.

The basic tool in the analysis is $U$-duality.  Type II supergravity
in $d$ dimensions has a moduli space of vacua of the form
$G/K$, where $G$ is a non-compact connected Lie group (which depends on
$d$) and $K$ is a compact
subgroup, generally a maximal compact subgroup of $G$.
$G$ is an exact symmetry of the supergravity theory.
There are also $U(1)$ gauge bosons,
whose charges transform as a representation of $G$.
\foot{To make a $G$-invariant theory on $G/K$, the matter fields
in general must be in representations of the unbroken symmetry
group $K$.  Matter fields that
are in representations of $K$ that do not extend to representations of $G$
are sections of some homogeneous vector bundles over $G/K$ with
non-zero curvature.  The potential existence of an integer lattice
of charges forces the gauge bosons to be sections instead of a flat
bundle, and that is why they are in a representation of $G$ and not only
of $K$.}
The structure
was originally found by dimensional reduction from eleven dimensions
\ref\julia{E. Cremmer and B. Julia, ``The $SO(8)$ Supergravity,''
Nucl. Phys. {\bf B159} (1979) 141; B. Julia, ``Group Disintegrations,'' in
{\it Superspace And Supergravity}, ed. M. Rocek and S. Hawking (Cambridge
University Press, 1981), p. 331.}.

In the string theory realization, the moduli space of vacua remains
$G/K$ since this is forced by the low energy supergravity.
Some of the Goldstone bosons parametrizing $G/K$ come from the NS-NS
sector and some from the RR sector.  The same is true of the gauge bosons.
In string theory, the gauge bosons that come from the NS-NS sector
couple to charged states in the elementary string spectrum.
It is therefore impossible for $G$ to be an exact symmetry of the string theory
-- it would not preserve the lattice of charges.  The $U$-duality conjecture
says that an integral form of $G$, call it $G({\bf Z})$, is a symmetry
of the string theory.  If so, then as the NS-NS gauge bosons couple
to BPS-saturated
charges, the same must be true of the RR gauge bosons -- though the
charges in question do not appear in the elementary string spectrum.
The existence of such RR charges was our main assumption in the last section;
we see that this assumption is essentially a consequence of $U$-duality.

The BPS saturated states are governed by an exact mass formula -- which
will be described later in some detail -- which shows how some of
them become massless when one approaches various limits in the moduli
space of vacua.  Our main dynamical assumption is that the smallest mass
scale appearing in the mass formula is always the smallest mass scale in
the theory.

We assume that at a generic point in $G/K$, the only massless states
are those in the supergravity multiplet.  There is then nothing to say
about the dynamics: the infrared behavior is that of $d$ dimensional Type II
supergravity.  There remains the question of what happens when one
takes various limits in $G/K$ -- for instance, limits that
correspond to weak coupling or large radius or (more mysteriously)
strong coupling or very strong excitation of RR scalars.
We will take the signal that something interesting is happening
to be that the mass formula predicts that some states are going to zero
mass.  When this occurs, we will try to determine what is the
dynamics of the light states, in whatever limit is under discussion.

We will get a complete answer, in the sense that
for every degeneration of the Type II
superstring in $d$ dimensions, there is a natural candidate for the
dynamics.  In fact, there are basically only two kinds of degeneration;
one involves weakly coupled string theory, and the other involves
weakly coupled
eleven-dimensional supergravity.  In one kind of degeneration, one
sees toroidal compactification of a Type II superstring from ten
to $d$ dimensions; the degeneration
consists of the fact that the string coupling constant is going to zero.
(The parameters of the torus are remaining fixed.)
In the other degeneration  one sees
toroidal compactification of eleven-dimensional supergravity from
eleven to $d$ dimensions; the degeneration
consists of the fact that  the radius of the torus is going to infinity
so that again the coupling constant at the compactification scale is going
to zero.\foot{It is only in the eleven-dimensional description that the
radius is going to infinity.  In the ten-dimensional string theory
description, the radius is fixed but the string coupling constant is
going to infinity.}
(These are actually the degenerations that produce maximal
sets of massless particles; others correspond to perturbations of these.)

Thus, with our hypotheses,
one gets a complete control on the dynamics, including strong coupling.
Every limit which one might have been tempted to describe
as ``strong coupling'' actually has a weakly coupled description
in the appropriate variables.
The ability to get this consistent picture can be taken as evidence
that the hypotheses are true, that $U$-duality is valid, and that
eleven-dimensional supergravity plays the role in the dynamics that
was claimed in section two.

It may seem unexpected that weakly coupled string theory appears in this
analysis as a ``degeneration,'' where some particles go to zero mass,
so let me explain this situation.
For $d<9$, $G$ is semi-simple, and the dilaton is unified with
other  scalars.  The ``string'' version of the low energy
effective action, in which the dilaton is singled out in the
gravitational kinetic energy
\eqn\yoyo{\int d^dx \sqrt g e^{-2\phi}R}
is unnatural for exhibiting such a symmetry.  The $G$-invariant
metric is the one obtained by a Weyl transformation that removes
the $e^{-2\phi}$ from the gravitational kinetic energy. The
transformation in question is of course the change of variables
$g=e^{4\phi/(d-2)}g'$, with $g'$ the new metric.
This transformation multiplies masses by $e^{2\phi/(d-2)}$,
that is, by
\eqn\kudos{w_d=\lambda^{2/(d-2)}}
(with $\lambda$ the string coupling
constant).  Thus, while elementary string states have masses
of order one with respect to the string metric, their masses are of
order $\lambda^{2/(d-2)}$ in the natural units for discussions
of $U$-duality.  So, from this point of view, the region
of weakly coupled string theory is a ``degeneration'' in which some masses
go to zero.

It is amusing to consider that, in a world in which supergravity was
known and string theory unknown, the following discussion might have been
carried out, with a view to determining the strong coupling limit of
a hypothethical consistent theory related to Type II supergravity.
The string theory degeneration might then have been found, giving a clue
to the existence of this theory.  Similarly, the strong coupling
analysis that we are about to perform might {\it a priori} have
uncovered new theories beyond string theory and eleven-dimensional
supergravity, but this will not be the case.

\subsec{The Nature Of Infinity}

It is useful to first explain -- without specific computations --
why NS-NS (rather than RR) moduli play the primary role.

We are interested in understanding what particles become light
-- and how they interact -- when one goes to infinity in the
moduli space $G({\bf Z})\backslash G/K$.  The discussion is simplified
by the fact that the groups $G$ that arise in supergravity are the
maximally split forms of the corresponding Lie groups.  This simply
means that they contain a maximal abelian subgroup $A$ which is a product
of copies  of ${\bf R}^*$ (rather than $U(1)$).
\foot{Algebraists call $A$ a ``maximal torus,'' and $T$ would be the
standard name, but I will avoid this terminology because
(i) calling $({\bf R}^*)^n$ a ``torus'' might be confusing, especially
 in the present
context in which there are so many other tori;
(ii) in the present problem the letter $T$ is reserved for
the $T$-duality group.}

For instance, in six dimensions $G=SO(5,5)$, with rank 5.
One can think of $G$ as the orthogonal group acting on the sum of
five copies of a two dimensional real vector space $H$
endowed with quadratic form
\eqn\endowed{\left(\matrix{ 0 & 1\cr 1 & 0 \cr}\right).}
Then a maximal abelian subgroup of $G$ is the space of matrices
looking like a sum of five $2\times 2$ blocks, of the form
\eqn\dwoed{\left(\matrix{ e^{\lambda_i} & 0\cr 0& e^{-\lambda_i}  \cr}\right)}
for some $\lambda_i$.  This group is of the form $({\bf R}^*)^5$.
Likewise, the integral forms arising in $T$ and $U$ duality are the maximally
split forms over ${\bf Z}$; for instance the $T$-duality group upon
compactification to $10-d$ dimensions is the group of integral matrices
preserving a quadratic form which is the sum of $d$ copies of \endowed.
This group is sometimes called $SO(d,d;{\bf Z})$.

With the understanding that
$G$ and $G({\bf Z})$ are the maximally split forms, the structure of
infinity in $G({\bf Z})\backslash G/K$ is particularly simple.
A fundamental domain in $G({\bf Z})\backslash G/K$ consists of group
elements of the form $g=tu$, where the notation is as follows.
$u$ runs over a {\it compact} subset $U$ of the space of generalized
upper triangular matrices; compactness of $U$ means that motion in
$U$ is irrelevant in classifying the possible ways to ``go to infinity.''
$t$ runs over $A/W$ where $A$ was described above,
 and $W$ is the Weyl group.

Thus, one can really only go to infinity in the $A$ direction,
and moreover, because of dividing by $W$, one only has to consider
going to infinity in a ``positive'' direction.

Actually, $A$ has a very simple physical interpretation.
Consider the special case of compactification from 10 to $10-d$
dimensions on an orthogonal product of circles ${\bf S}^1_i$ of
radius $r_i$.  Then $G$ has rank $d+1$, so $A$ is a product of
$d+1$ ${\bf R}^*$'s.  $d$ copies of ${\bf R}^*$ act by rescaling the
$r_i$ (making up a maximal abelian subgroup of
the $T$-duality group $SO(d,d)$),
and the last one rescales the string
coupling constant.  So in particular, with this choice of $A$, if one
starts at a point in moduli space at which the RR fields are all zero,
they remain zero under the action of $A$.

Thus, one can probe all
possible directions at infinity without exciting the RR fields; directions
in which some RR fields go to infinity are equivalent to directions
in which one only goes to infinity via NS-NS fields.
Moreover, by the description of $A$ just given, going to infinity in
NS-NS directions can be understood to mean just taking the string
coupling constant and the radial parameters
of the compactification to zero or infinity.

\subsec{The Central Charges And Their Role}

Let us now review precisely why it is possible to
predict particle masses from $U$-duality.
The unbroken subgroup $K$ of the supergravity symmetry group
$G$ is realized in Type II supergravity as an $R$-symmetry group;
that is, it acts non-trivially on the supersymmetries.  $K$ therefore
acts on the central charges in the supersymmetry algebra.
The scalar fields parametrizing
the coset space $G/K$ enable one to write a $G$-invariant formula
for the central charges (which are a representation of $K$)
of the gauge bosons (which are a representation of $G$).  For most
values of $d$, the formula is uniquely determined, up to a multiplicative
constant, by $G$-invariance, so the analysis does not require
many details of supergravity.  That is fortunate as not all the details
we need have been worked out in the literature, though many can be found
in \ref\divers{A. Salam and E. Sezgin, eds., {\it
Supergravity In Diverse
Dimensions} (North-Holland/World Scientific, 1989).}.

For example, let us recall (following \hull) the situation in $d=4$.
The $T$-duality group is $SO(6,6)$, and $S$-duality would be $SL(2)$
(acting on the axion-dilaton system and exchanging electric and magnetic
charge).  $SO(6,6)\times SL(2)$ is a maximal subgroup of the $U$-duality group
which is $G=E_7$ (in its non-compact, maximally split form)
and has $K=SU(8)$ as a maximal compact subgroup.

Toroidal compactification from ten
to four dimensions produces  in the NS-NS sector
twelve gauge bosons coupling to string momentum and winding states,
and transforming in the twelve-dimensional
representation of $SO(6,6)$.
The electric and magnetic charges coupling to any one of these
gauge bosons transform
as a doublet of $SL(2)$, so altogether the NS-NS sector generates a total
of 24 gauge charges, transforming as $({\bf 12},{\bf 2})$
of $SO(6,6)\times SL(2)$.

{}From the RR sector, meanwhile, one gets 16 vectors.  (For instance,
in Type IIA, the vector of the ten-dimensional RR sector gives
1 vector in four dimensions, and the three-form gives $6\cdot 5/2=15$.)
These 16 states give  a total of $16\cdot 2= 32$
electric and magnetic charges,
which can be argued to transform in an irreducible
spinor representation of $SO(6,6)$ (of positive or negative chirality for
Type IIA or Type IIB), while being $SL(2)$ singlet.
The fact that these states are $SL(2)$ singlets means that there is no
natural way to say which of the RR charges are electric and which are
magnetic. Altogether, there are $24+32=56$ gauge charges,
transforming as
\eqn\ilnco{({\bf 12},{\bf 2})\oplus ({\bf 32},{\bf 1})}
under $SO(6,6)\times SL(2)$; this is the decomposition of the irreducible
${\bf 56}$ of $E_7$.  Let us call the space of
these charges $V$.

The four-dimensional theory has $N=8$ supersymmetry; thus
there are eight positive-chirality supercharges $Q_\alpha^i$,
$i=1\dots 8$, transforming in the ${\bf 8}$ of $K=SU(8)$.
The central charges, arising in the formula
\eqn\hutty{\{Q_\alpha^i,Q_\beta^j\}=\epsilon_{\alpha\beta}Z^{ij},}
therefore transform as the second rank antisymmetric tensor of $SU(8)$,
the ${\bf 28}$: this representation has complex dimension 28 or real
dimension 56.  Denote the space of $Z^{ij}$'s as $W$.

Indeed, the ${\bf 56}$ of $E_7$, when restricted
to $SU(8)$, coincides with the ${\bf 28}$, regarded as a 56-dimensional
real representation.  (Equivalently, the ${\bf 56}$ of $E_7$ when
complexified decomposes as ${\bf 28}\oplus \bar{\bf 28}$ of $SU(8)$.)
There is of course a natural, $SU(8)$-invariant metric on $W$.
As the ${\bf 56}$ is a pseudoreal rather than real representation of $E_7$,
there is no $E_7$-invariant metric on $V$.  However, as $V$ and $W$
coincide when regarded as representations of $SU(8)$, one can pick
an embedding of $SU(8)$ in $E_7$ and then define an $SU(8)$-covariant
map $T:V\to W$ which determines a metric on $V$.

There is no reason to pick one embedding rather than another, and indeed
the space of vacua $E_7/SU(8)$ of the low energy supergravity theory
can be interpreted as the space of all $SU(8)$ subgroups of $E_7$.
Given $g\in E_7$, we can replace $T:V\to W$
by
\eqn\jury{T_g=Tg^{-1}.}
This is not invariant under $g\to gk$, with $k\in SU(8)$, but it
is so invariant up to an $SU(8)$ transformation of $W$.
So let $\psi\in V$ be a vector of gauge charges of some string state.
Then
\eqn\pury{\psi \to Z(\psi)=T_g\psi}
gives a vector in $W$, representing the central charges
of $\psi$.  The map from ``states'' $\psi$ to central charges $Z(\psi)$
 is manifestly $E_7$-invariant, that is invariant
under
\eqn\lury{\eqalign{\psi & \to g'\psi\cr
                    g   & \to g' g. \cr}}
Also, under $g\to gk$, with $k\in SU(8)$, $Z$ transforms to $Tk^{-1}T^{-1}Z$,
that is, it transforms by a ``local $SU(8)$ transformation'' that does not
affect the norm of the central charge.  The formula \pury\ is, up
to a constant multiple, the only formula with these properties, so
it is the one that must come from the supergravity or superstring theory.

In supersymmetric theories with central charges, there is an
inequality between the mass of a state
and the central charge.
For elementary string winding states and their partners under
$U$-duality, the inequality is $M\geq |Z|$.
(More generally, the inequality is roughly that $M$ is equal to or
greater than the
largest eigenvalue of $Z$; for a description of stringy black
holes with more than one eigenvalue, see \strathdee. Elementary string
states have only one eigenvalue.)

So far, we have not mentioned the integrality of the gauge charges.
Actually, states carrying the 56 gauge charges
 only populate a lattice
$V_{\bf Z}\subset V$.   If $U$-duality is true, then each
lattice point related by $U$-duality to the gauge charges of an elementary
string state
represents the charges of a supermultiplet
of mass $|Z(\psi)|$.

As an example of the use of this formalism, let us keep a promise
made in section two  and give an alternative deduction, assuming $U$-duality,
of the important statement that the masses of states carrying RR charges
are (in string units) of order $1/\lambda$.\foot{The following argument
was pointed out in parallel by C. Hull.}  Starting from any given
vacuum, consider the one-parameter family of vacua determined by the
following one-parameter subgroup of $SO(6,6)\times SL(2)$:
we take the identity in $SO(6,6)$ (so that the parameters of the toroidal
compactification are constant) times
\eqn\imco{g_t=\left(\matrix{e^{t } & 0 \cr 0 & e^{-t} \cr}\right)}
in $SL(2)$ (so as to vary the string coupling constant).
We work here in a basis in which the ``top'' component
is electric and the ``bottom'' component is magnetic.

Using the mass formula $M(\psi)=|Z(\psi)|=|Tg^{-1}\psi|$,
the $t$ dependence of
the mass of a state comes entirely from the $g$ action on the state.
The NS-NS states, as they are in a doublet of $SL(2)$, have
``electric'' components whose masses scale as $e^{-t}$ and ``magnetic''
components with masses of $e^t$.  On the other hand, as the RR states
are $SL(2)$ singlets, the mass formula immediately implies that their
masses are independent of $t$.

These are really the masses in the $U$-dual ``Einstein'' metric.
Making a Weyl transformation to the ``string'' basis  in which
the electric NS-NS states (which are elementary string states)
have masses of order one, the masses are as follows:
electric NS-NS, $M\sim 1$; magnetic NS-NS, $M\sim e^{2t}$; RR,
$M\sim e^t$.  But since we know that the magnetic NS-NS states
(being fairly conventional solitons)
have masses of order $1/\lambda^2$, we identify $e^t=1/\lambda$
(a formula one could also get from the low energy
supergravity); hence the RR masses are of order $1/\lambda$ as claimed.
\foot{We made this deduction here in four dimensions, but it could be made,
using $U$-duality, in other dimensions as well.  Outside of four
dimensions, instead of using the known mass scale of magnetic monopoles
to fix the relation between $t$ and $\lambda$, one could use the known
Weyl transformation \kudos\ between the string and $U$-dual mass scales.}

The basic properties described above hold in any dimension above three.
(In nine dimensions, some extra care is needed because the $U$-duality
group is not semi-simple.)
In three dimensions, new phenomena, which we will not try to
unravel, appear because vectors are dual to scalars and charges are confined
(for some of the relevant material, see \ref\threesen{A. Sen,
``Strong-Weak Coupling Duality In Three-Dimensional String Theory,''
hepth/9408083.}).

\subsec{Analysis Of Dynamics}

We now want to justify the claims made at the beginning of this section
about the strong coupling dynamics.

To do this, we will
analyze limits of the theory in which some of the BPS-saturated particles
go to zero mass.  Actually,
for each way of going to infinity, we will look only at the particles
whose masses goes to zero as fast as possible.
We will loosely call these the particles that are massless at infinity.

Also, we really want to find the ``maximal'' degenerations,
which produce maximal
sets of such massless particles; a set of massless particles, produced
by going to infinity in some direction, is maximal
if there would be no way of going to infinity such that those particles
would become massless together with others.  A degeneration
(i.e., a path to infinity) that produces a non-maximal set of
massless particles should be understood as a perturbation of
a maximal degeneration.
(In field theory, such perturbations, which partly lift the degeneracy
of the massless particles, are called perturbations by relevant operators.)
We will actually also check a few non-maximal degenerations, just
to make sure that we understand their physical interpretation.

To justify our claims, we should show that in any dimension $d$, there
are only two maximal degenerations, which correspond to toroidal
compactification of weakly coupled ten-dimensional string theory
and to toroidal compactification of eleven-dimensional supergravity,
respectively.  The analysis is in fact very similar in spirit for any
$d$, but the details of the group theory are easier for some values of
$d$ than others.  I will first explain a very explicit analysis
for $d=7$, chosen as a relatively easy case, and then explain
an efficient approach for arbitrary $d$.

In $d=7$, the $T$-duality group is $SO(3,3)$, which is the same as
$SL(4)$; $U$-duality extends this to $G=SL(5)$.  A maximal compact
subgroup is $K=SO(5)$.

In the NS-NS sector, there are six $U(1)$ gauge fields
that come from the compactification
on a three-torus; they transform as a vector of $SO(3,3)$ or second
rank antisymmetric tensor of $SL(4)$.  In addition; four more $U(1)$'s,
transforming as a spinor of $SO(3,3)$ or a ${\bf 4}$ of $SL(4)$,
come from the RR sector.  These states combine with the six from the NS-NS
sector to make the second rank antisymmetric tensor, the ${\bf 10}$ of
$SL(5)$.

In Type II supergravity in seven dimensions, the maximal possible
$R$-symmetry is $K=SO(5)$ or $Sp(4)$.  The supercharges make up in fact
four pseudo-real spinors $Q_\alpha^i$, $i=1\dots 4$, of the seven-dimensional
Lorentz group $SO(1,6)$,
transforming as the ${\bf 4}$ of $Sp(4)$. The central charges transform
in the symmetric part of ${\bf 4}\times {\bf 4}$, which is the
${\bf 10}$ or antisymmetric tensor of $SO(5)$.  Thus, we are in
a situation similar to what was described earlier in four dimensions:
the gauge charges transform as the ${\bf 10}$ of
$SL(5)$, the central charges transform in the ${\bf 10}$ of $SO(5)$,
and a choice of vacuum in $G/K=SL(5)/SO(5)$ selects an $SO(5)$ subgroup
of $SL(5)$, enabling one to identify these representations and map
gauge charges to central charges.

A maximal abelian subgroup $A$ of $SL(5)$ is given by the diagonal matrices.
A one-parameter subgroup of $A$ consists of matrices of the form
\eqn\jippo{g_t=\left(\matrix{e^{a_1t} & 0 & 0 & 0 &  0 \cr
                                0  & e^{a_2t} & 0 & 0 & 0 \cr
                                0  &  0  &   e^{a_3t} & 0 & 0 \cr
                                0  &  0  &    0    & e^{a_4t} & 0 \cr
                                0  &  0  &    0    &   0   & e^{a_5t}\cr}
        \right)}
where the $a_i$ are constants, not all zero,
with $\sum_i a_i=0$.  We want to consider the behavior of the spectrum
as $t\to +\infty$. By a Weyl transformation,
we can limit ourselves to the case that
\eqn\kippo{a_1\geq a_2\geq \dots \geq a_5.}

Let $\psi_{ij}$, $i<j$ be a vector in the ${\bf 10}$ of $SL(5)$ whose
components are zero except for the $ij$ component, which is 1
(and the $ji$ component, which is $-1$).  We will also use the name
$\psi_{ij}$ for a particle with those gauge charges.
The mass formula $M(\psi)=|Tg^{-1}\psi|$
says that the mass of $\psi_{ij}$ scales with $t$ as
\eqn\anson{M(\psi_{ij})\sim e^{-t(a_i+a_j)}.}
By virtue of \kippo, the lightest type of particle
is $\psi_{12}$.  For generic values of the $a_i$,
this is the unique particle whose mass scales
to zero fastest, but if $a_2=a_3$ then $\psi_{12}$ is degenerate
with other particles.  To get a maximal set of particles
degenerate with $\psi_{12}$, we need a maximal set of $a_i$ equal
to $a_2$ and $a_3$.  We cannot set all $a_i$ equal (then they have
to vanish, as $\sum_i a_i=0$),
so by virtue of \kippo, there are two maximal cases, with
$a_1=a_2=a_3=a_4$, or $a_2=a_3=a_4=a_5$.  So the maximal degenerations
correspond to one-parameter subgroups
\eqn\jjippo{g_t=\left(\matrix{e^{t} & 0 & 0 & 0 &  0 \cr
                                0  & e^{t} & 0 & 0 & 0 \cr
                                0  &  0  &   e^{t} & 0 & 0 \cr
                                0  &  0  &    0    & e^{t} & 0 \cr
                                0  &  0  &    0    &   0   & e^{-4t}\cr}
        \right)}
or
\eqn\jjjippo{g_t=\left(\matrix{e^{4t} & 0 & 0 & 0 &  0 \cr
                                0  & e^{-t} & 0 & 0 & 0 \cr
                                0  &  0  &   e^{-t} & 0 & 0 \cr
                                0  &  0  &    0    & e^{-t} & 0 \cr
                                0  &  0  &    0    &   0   & e^{-t}\cr}
        \right)}
with $t\to +\infty$.  As we will see, the first corresponds to weakly
coupled string theory, and the second to eleven-dimensional supergravity.

In \jjippo, the particles whose masses vanish for $t\to +\infty$
are the $\psi_{ij}$ with $1\leq i<j
\leq 4$.  There are six of these, the correct number of light elementary
string states of string theory compactified from ten to seven dimensions.
Moreover, in \jjippo, $g_t$ commutes with a copy of $SL(4)$ that
acts on indices $1-2-3-4$.  This part of the seven-dimensional
symmetry group $SL(5)$ is unbroken by going to infinity in the direction
\jjippo, and hence would be observed as a symmetry of the low energy
physics at ``infinity'' (though most of the symmetry is spontaneously
broken in any given vacuum near infinity).
Indeed, $SL(4)$ with six gauge charges in the antisymmetric tensor
representation is the correct $T$-duality group of weakly coupled
string theory in seven dimensions.

There is a point here that may be puzzling at first sight.
The full subgroup of $SL(5)$
that commutes with $g_t$ is actually not $SL(4)$ but
$SL(4)\times {\bf R}^*$, where
${\bf R}^*$ is the one-parameter subgroup containing $g_t$.  What
happens to the ${\bf R}^*$?
When one restricts to the {\it integral} points in $SL(5)$, which are the
true string symmetries, this ${\bf R}^*$ does not contribute, so the
symmetry group at infinity is just the integral form of $SL(4)$.  A similar
comment applies at several points below and will not be repeated.

Moving on now to the second case,
in \jjjippo, the particles whose masses vanish for $t\to +\infty$
are the $\psi_{1 i}$, $i>1$.  There are four of these, the correct
number for compactification of eleven-dimensional supergravity on a
four-torus ${\bf T}^4$ whose dimensions are growing with $t$.
The gauge charges of light states are simply the components of the momentum
along ${\bf T}^4$.
The symmetry
group at infinity is again $SL(4)$.  This $SL(4)$ has a natural
interpretation as a group of linear
automorphisms of ${\bf T}^4$.
\foot{That is,
if ${\bf T}^4$ is understood as the space of real variables $y^i,$ $i=1\dots
4$, modulo $y^i\to y^i+n^i$, with $n^i\in {\bf Z}$, then $SL(4)$ acts
by $y^i\to w^i{}_jy^j$.  For this to be a diffeomorphism and preserve
the orientation, the determinant
of $w$ must be one, so one is in $SL(4)$.  Given an $n$-torus ${\bf T}^n$,
we will subsequently use the phrase ``mapping class group'' to refer
to the $SL(n)$ that acts linearly in this sense on ${\bf T}^n$.}
In fact, the gauge
charges carried by the light states in \jjjippo\ transform in the
${\bf 4}$ of $SL(4)$, which agrees with the supergravity description
as that is how the momentum components along ${\bf T}^4$ transform
under $SL(4)$.  As this $SL(4)$ mixes three of the ``original'' ten
dimensions with the eleventh dimension that is associated with strong
coupling, we have our first evidence for the underlying eleven-dimensional
Lorentz invariance.

Finally, let us consider a few non-maximal degenerations, to make
sure we understand how to interpret them.
\foot{We will see in the next section that when the $U$-duality group
has rank $r$, there are $r$ naturally distinguished one-parameter subgroups.
For $SL(5)$, these are \jjippo, \jjjippo, and the two introduced below.}
Degeneration in the direction
\eqn\jujippo{g_t=\left(\matrix{e^{3t} & 0 & 0 & 0 &  0 \cr
                                0  & e^{3t} & 0 & 0 & 0 \cr
                                0  &  0  &   e^{-2t} & 0 & 0 \cr
                                0  &  0  &    0    & e^{-2t} & 0 \cr
                                0  &  0  &    0    &   0   & e^{-2t}\cr}
        \right)}
leaves as $t\to\infty$
the unique lightest state $\psi_{12}$.  I interpret this as coming
from partial decompactification to eight dimensions
-- taking one circle much larger than the others so that the elementary
string states with momentum in that one direction are the lightest.
This family has the symmetry group $SL(3)\times SL(2)$, which is indeed
the $U$-duality group in eight dimensions, as it should be.

The family
\eqn\ujjippo{g_t=\left(\matrix{e^{2t} & 0 & 0 & 0 &  0 \cr
                                0  & e^{2t} & 0 & 0 & 0 \cr
                                0  &  0  &   e^{2t} & 0 & 0 \cr
                                0  &  0  &    0    & e^{-3t} & 0 \cr
                                0  &  0  &    0    &   0   & e^{-3t}\cr}
        \right)}
gives three massless states $\psi_{ij}$, $1\leq i<j\leq 3$, transforming
as $({\bf 3},{\bf  1})$ of the symmetry group $SL(3)\times SL(2)$.
I interpret this as decompactification to the Type IIB theory in
ten dimensions -- taking all three circles to be very large.
The three light charges are the momenta around the three circles;
$SL(3)$ is the mapping class group of the large three-torus, and
$SL(2)$ is the $U$-duality group of the Type IIB theory in ten dimensions.

\bigskip
\noindent{\it Partially Saturated States}

I will now justify an assumption made above and also make a further
test of the interpretation that we have proposed.

First of all, we identified BPS-saturated
elementary string states with charge tensors
$\psi_{ij}$ with (in the right basis) only one non-zero entry.
Why was this valid?

We may as well consider NS-NS states; then we can restrict ourselves
to the $T$-duality group $SO(3,3)$.  The gauge charges transform in
the vector representation of $SO(3,3)$.
Given such a vector $v_a$, one can define the quadratic
invariant $(v,v)=\sum_{a,b}\eta^{ab}v_av_b$.

On the other hand, $SO(3,3)$ is the same as $SL(4)$, and $v$ is equivalent
to a second rank antisymmetric tensor $\psi $ of $SL(4)$.
In terms of $\psi$, the quadratic invariant is $(\psi,\psi)={1\over 4}
\epsilon^{ijkl}\psi_{ij}\psi_{kl}$.
By an $SL(4)$ transformation, one can bring $\psi $ to a normal
form in which the independent
non-zero entries are $\psi_{12}$ and $\psi_{34}$ only.  Then
\eqn\shns{(\psi,\psi)=2\psi_{12}\psi_{34}.}
So the condition that the particle carries only one type of charge,
that is, that only $\psi_{12}$ or $\psi_{34}$ is non-zero, is that
$(\psi,\psi)=0$.

Now let us consider the elementary string states.  Such a state
has in the toroidal directions left- and right-moving momenta
$p_L$ and $p_R$.  $p_L$ and $p_R$ together form a vector of $SO(3,3)$,
and the quadratic invariant is \narain\
\eqn\nonesen{(p,p)=|p_L|^2-|p_R|^2.}
BPS-saturated states have no oscillator excitations for left- or right-movers,
and the mass shell condition requires that they obey $|p_L|^2-|p_R|^2=0$,
that is, that the momentum or charge vector $p$ is light-like.
This implies, according to the discussion in the last paragraph, that
in the right basis, the charge tensor $\psi$ has only one entry.
That is the assumption we made.

Now, however, we can do somewhat better and consider elementary
string states of Type II that are BPS-saturated for left-movers
only (or equivalently, for right-movers only).  Such states are
in ``middle-sized'' supermultiplets, of dimension $2^{12}$
(as opposed to generic supermultiplets of dimension $2^{16}$ and
BPS-saturated multiplets of dimension $2^8$).  To achieve BPS saturation
for the left-movers only, one puts the left-moving oscillators in their
ground state, but one permits right-moving oscillator excitations;
as those excitations are arbitrary, one gets an exponential spectrum
of these half-saturated states (analogous to the exponential spectrum of
BPS-saturated states in the heterotic string \ref\harvey{A. Dabholkar
and J. A. Harvey, ``Nonrenormalization Of The Superstring Tension,''
Phys. Rev. Lett. {\bf 63} (1989) 719;
A. Dabholkar, G. Gibbons, J. A. Harvey, and F. Ruiz Ruiz, ``Superstrings
And Solitons,''
Nucl. Phys. {\bf B340} (1990) 33.}).
With oscillator excitations for right-movers only, the mass shell condition
implies that $|p_L|^2>|p_R|^2$, and hence the charge vector is not lightlike.
The charge tensor $\psi$ therefore in its normal form
has both $\psi_{12}$ and $\psi_{34}$ non-zero.  For such states,
the mass inequality says that the mass is bounded below by the largest
eigenvalue of $Tg^{-1}\psi$, with
equality for the ``middle-sized'' multiplets.

With this in mind, let us consider the behavior of such half-saturated
states in the various degenerations.
In the ``stringy'' degeneration \jjippo, a state with non-zero
$\psi_{12}$ and $\psi_{34}$ has a mass of the same order of magnitude
as a state with only $\psi_{12}$ non-zero.  This is as we would
expect from weakly coupled string theory with toroidal radii of order one:
the half-saturated states
have masses of the same order of magnitude as the BPS-saturated massive
modes.  To this extent, string excitations show up in the strong coupling
analysis.

What about the ``eleven-dimensional'' degeneration \jjjippo?
In this case, while the particles with only one type of charge have
masses that vanish as $e^{-3t}$ for $t\to\infty$, the particles
with two kinds of charge have masses that {\it grow} as $e^{+t}$.
The only light states that we can see with this formalism
in this degeneration are the Kaluza-Klein modes of eleven-dimensional
supergravity.  There is, for instance,  no evidence for membrane excitations;
such evidence might well have appeared if a consistent membrane theory
with eleven-dimensional supergravity as its low energy limit really
does exist.

\subsec{Framework For General Analysis}

It would be tiresome to repeat this analysis ``by hand'' in other
values of the dimension.  Instead,  I will now\foot{With some assistance
from A. Borel.} explain a bit of group theory that makes the analysis
easy.  One of the main points is to incorporate the action of the Weyl
group. This was done above by choosing $a_1\geq a_2\geq \dots \geq a_5$, but
to exploit the analogous condition in $E_7$, for instance, a little
machinery is useful.

In $d$ dimensions, the $U$-duality group $G$ has rank $r=11-d$.
Given any one-parameter subgroup $F$  of a maximal abelian subgroup $A$,
one can pick a set of simple positive roots $x_i$ such that the
action of $F$ on the $x_i$ is
\eqn\jsons{x_i\to e^{c_it}x_i}
with $c_i$ {\it non-negative}.  In this restriction on the $c_i$, we have
used the Weyl action.  Conversely, for every set of non-negative $c_i$
(not all zero), there is a one-parameter subgroup $F$ that acts as \jsons.

The gauge charges are in some representation $R$ of $G$; that is, for each
weight in $R$ there is a corresponding gauge charge.
\foot{The particular representations $R$ that actually arise in Type
II string theory in $d\geq 4$
 have the
property (unusual among representations of Lie groups) that the
non-zero weight spaces are all one-dimensional.  It therefore
makes sense to label the gauge charges by weights. (These representations
are actually ``minuscule'' -- the Weyl group acts transitively on the
weights.)   $d\leq 3$ would
have some new features, as already mentioned above.}
Let $\rho=
\sum_ie_ix_i$ be the highest weight in $R$.  The $e_i$ are positive
integers.
A particle whose only gauge charge is the one that corresponds to $\rho$
has a mass that vanishes for $t\to +\infty$ as
\eqn\juppe{M_\rho\sim \exp\left(-\sum_i c_ie_i t\right).}
Any other weight in $R$ is of the form $\rho'=\sum_if_ix_i$,
with $f_i\leq e_i$.  A particle carrying the $\rho'$ charge has mass
of order
\eqn\uppe{M_{\rho'}\sim \exp\left(-\sum_ic_if_it\right).}
Thus $M_{\rho'}\geq M_\rho$ -- the particle with only charge $\rho$
always goes to zero mass at least as fast as any other -- and $M_{\rho'}=
M_\rho$ if and only if
\eqn\hupep{c_i=0~{\rm whenever}~f_i<e_i.}

Now, our problem is to pick the subgroup $F$, that is, the $c_i$, so
that a maximal set of $M_{\rho'}$ are equal to $M_\rho$.  If the $c_i$
are all non-zero, then (as the highest weight state is unique) \hupep\
implies that $\rho'=\rho$ and only one gauge charge is carried by the lightest
particles.  The condition in \hupep\ becomes less restrictive only when
one of the $c_i$ becomes zero, and to get a maximal set of  $M_{\rho \, '}$
degenerate with $M_\rho$, we must set as many of the $c_i$ as possible
to zero.  As the $c_i$ may not all vanish, the best we can do is
to set $r-1$ of them to zero.  There are therefore precisely $r$ one-parameter
subgroups $F_i$ to consider, labeled by which of the $c_i$ is non-zero.

The $x_i$ are labeled by the vertices in the Dynkin diagram of $G$,
so each of the $F_i$ is associated with a particular vertex $P_i$.
Deleting $P_i$ from the Dynkin diagram of $G$ leaves the Dynkin diagram
of a rank $r-1$ subgroup $H_i$ of $G$.  It is the unbroken subgroup
when one goes to infinity in the $F_i$ direction.

\subsec{Analysis In $d=4$}

With this machinery, it is straightforward to analyze the dynamics
in each dimension $d$. As the rank is $r=11-d$,
there are $11-d$ distinguished one-parameter
subgroups to check.  It turns out that one of them corresponds to weakly
coupled string theory in $d$ dimensions, one to toroidal compactification
of eleven-dimensional supergravity to $d$ dimensions,
and the others to partial (or complete) decompactifications.
In each case, the symmetry group when one goes to infinity is the
expected one: the $T$-duality group $SO(10-d,10-d)$ for the string
degeneration; the mapping class group $SL(11-d)$ for supergravity;
or for partial decompactification to $d'$ dimensions, the product
of the mapping class group $SL(d'-d)$ of a $d'-d$-torus and the
$U$-duality group in $d'$ dimensions.

I will illustrate all this
in $d=4$, where the $U$-duality group is $E_7$.  Going to infinity
in a direction $F_i$ associated with one of the seven points in the Dynkin
diagram leaves as unbroken subgroup  $H_i$ one of the following:

(1) $SO(6,6)$: this is the $T$-duality group for string theory
toroidally compactified from ten to four dimensions.
This is a maximal degeneration, with (as we will see) 12 massless states
transforming in the ${\bf 12}$ of $SO(6,6)$.

(2) $SL(7)$: this is associated with eleven-dimensional supergravity
compactified to four dimensions on a seven-torus whose mapping
class group is $SL(7)$.  This is the other
maximal degeneration; there are the expected seven massless states in the
${\bf 7}$ of $SL(7)$.

(3) $E_6$: this and the other cases are non-maximal degenerations
corresponding to partial decompactification.  This case corresponds
to partial decompactification to five dimensions by taking
one circle to be much larger than the others; there is only one massless
state, corresponding to a state with momentum around the large circle.
$E_6$ arises as the $U$-duality group in five dimensions.

(4) $SL_2\times SO(5,5)$: this is associated with partial decompactification
to six dimensions.  There are two light states, corresponding to momenta
around the two large circles; they transform as $({\bf 2},{\bf 1})$
under $SL_2\times SO(5,5)$.  $SL_2$ acts on the two large circles and
$SO(5,5) $ is the $U$-duality group in six dimensions.

(5) $SL_3\times SL(5)$: this is associated with partial decompactification
to seven  dimensions.  $SL(3)$ acts on the three large circles (and the
three light charges), and
$SL(5)$ is the $U$-duality in seven  dimensions.

(6) $SL_4\times SL(3)\times SL(2)$:
this is associated with partial decompactification
to eight dimensions.  $SL(4)$ acts on the four large circles and light
charges, and $SL(3)\times SL(2) $ is $U$-duality in eight dimensions.

(7) $SL_6\times SL_2$: this is associated with
decompactification to Type IIB in ten dimensions.  $SL_6$ acts on the six large
circles and light charges, and $SL(2)$ is the $U$-duality in
ten dimensions.

In what follows, I will just check the assertions
about the light spectrum
for the first two cases, which are the important ones, and the third,
which is representative of the others.

(1) $F_1$ can be described as follows.  $E_7$ contains a
maximal subgroup $SO(6,6)\times SL(2)$.
$F_1$ can be taken as
the subgroup of $SL(2)$ consisting of matrices of the form
\eqn\hging{\left(\matrix{e^t & 0 \cr 0 & e^{-t} \cr}\right).}
The gauge charges are in the ${\bf 56}$ of $E_7$, which decomposes
under $L_1$ as $({\bf 12},{\bf 2})\oplus ({\bf 32},{\bf 1})$.
The lightest states come from the part of the $({\bf 12},{\bf 2})$
that transforms as $e^t$ under \hging; these are the expected twelve
states in the ${\bf 12}$ of $SO(6,6)$.

(2) $E_7$ contains a maximal subgroup $SL(8)$.
$F_2$ can be taken as the subgroup of $SL(8)$ consisting of group elements
$g_t={\rm diag}(e^t,e^t,\dots,e^t,e^{-7t})$.
The ${\bf 56}$ of $E_7$ decomposes as ${\bf 28}\oplus
{\bf 28}'$ -- the antisymmetric tensor plus its dual.
The states of highest eigenvalue (namely $e^{8t}$)
are seven states in the ${\bf 28}$  transforming in the expected
${\bf 7}$ of the unbroken $SL(7)$.

(3) $E_7$ has a maximal subgroup $E_6\times {\bf R}^*$, and $F_3$
is just the ${\bf R}^*$.
The ${\bf 56}$ of $E_7$ decomposes as ${\bf 27}^1\oplus {{\bf  27}'}^{-1}
\oplus {\bf 1}^{3}\oplus {\bf 1}^{-3}$, where the $E_6$ representation
is shown in boldface and the ${\bf R}^*$ charge (with some normalization)
by the exponent.  Thus in the $F_3$ degeneration, there is a unique
lightest state, the ${\bf 1}^3$.

The reader can similarly analyze the light spectrum for the other $F_i$,
or the analogous subgroups in $d\not= 4$.

\newsec{Heterotic String Dynamics Above Four Dimensions}

\subsec{A Puzzle In Five Dimensions}

$S$-duality gives an attractive proposal for the strong coupling
dynamics of the heterotic string after toroidal compactification to four
dimensions: it is equivalent to the same theory at weak coupling.
In the remainder of this paper,
we will try to guess the behavior above four dimensions.
This process will also yield some new insight about $S$-duality
in four dimensions.

Toroidal compactification of the heterotic string from 10 to $d$ dimensions
gives $2(10-d)$ vectors that arise from dimensional reduction of the
metric and antisymmetric tensor.  Some of the elementary string states
are electrically charged with respect to these vectors.

Precisely in five dimensions, one more vector arises.  This is so because
in five dimensions a two-form $B_{mn}$ is dual to a vector $A_m$,
roughly by $dB=*dA$.  In the elementary string spectrum, there are no
particles that are electrically charged with respect to $A$, roughly
because $A$ can be defined (as a vector) only in five dimensions.
But it is easy to see where to find such electric charges.  Letting
$H$ be the field strength of $B$ (including the Chern-Simons terms)
the anomaly equation
\eqn\murky{dH=\tr F\wedge F -\tr R\wedge R }
($F$ is the $E_8\times E_8$ or $SO(32)$
field strength and $R$ the Riemann tensor)
implies that the electric current of $A$ is
\eqn\urky{J=*\tr F\wedge F - *\tr R\wedge R.}
Thus, with $G=dA$, \murky\ becomes
\eqn\burky{D^mG_{mn}=J_n,}
showing that $J_n$ is the electric current.
So the charge density $J_0$ is the instanton density,
and a Yang-Mills instanton, regarded as a soliton in $4+1$
dimensions, is electrically charged with respect to $A$.

\nref\strominger{A. Strominger, ``Heterotic Solitons,'' Nucl. Phys.
{\bf B343} (1990) 167.}
\nref\callan{C. G. Callan, Jr., J. A. Harvey, and A. Strominger,
``World-Sheet Approach To Heterotic Instantons And Solitons,''
Nucl. Phys. {\bf B359} (1991) 611.}
Instantons (and their generalizations to include the supergravity
multiplet \refs{\strominger,\callan}) are invariant under one half
of the supersymmetries.  One would therefore suspect that
quantization of the instanton would give BPS-saturated multiplets,
with masses given by the instanton action:
\eqn\ucco{M= {16\pi^2|n|\over\lambda^2}.}
Here $n$ is the instanton number or electric charge and $\lambda$ is
the string coupling.

To really prove existence of these multiplets, one would
need to understand and quantize the collective coordinates of the stringy
instanton.
In doing this, one needs to pick a particular vacuum to work in.
In the generic toroidal vacuum, the unbroken gauge group is just a product
of $U(1)$'s.  Then the instantons, which require a non-abelian structure,
tend to shrink to zero size, where stringy effects are strong and
the analysis is difficult.  Alternatively, one can consider a special
vacuum with an unbroken non-abelian group, but this merely adds infrared
problems to the stringy problems.  The situation is analogous to the
study \ref\gaunt{J. Gauntlett and J. A. Harvey ``$S$-Duality And The
Spectrum Of Magnetic Monopoles In Heterotic String Theory,''
hepth/9407111.} of $H$-monopoles
after toroidal compactification to four dimensions; indeed, the present
paper originated with an effort to resolve the problems concerning
$H$-monopoles.  (The connection between instantons and $H$-monopoles is simply
that upon compactification of one of the spatial directions on a circle,
the instantons become what have been called $H$-monopoles.)

Despite the difficulty in the collective coordinate
analysis, there are two good reasons to believe
that BPS-saturated multiplets in this sector do exist.  One, already
mentioned, is the invariance of the classical solution under half
the supersymmetries. The second reason is that if in five dimensions,
the electrically charged states had masses bounded strictly above the
BPS value in \ucco, the same would be true after compactification
on a sufficiently big circle, and then the BPS-saturated $H$-monopoles
required for $S$-duality could not exist.

Accepting this assumption, we are in a similar situation to that
encountered earlier for
the Type IIA string in ten dimensions: there is a massless
vector, which couples to electric charges whose mass diverges for
weak coupling.  (The mass is
here proportional to $1/\lambda^2$ in contrast to $1/\lambda$ in
the other case.)
Just as in the previous situation, we have a severe puzzle if we take
the formula seriously for strong coupling, when these particles seem
to go to zero mass.

If we are willing to take \ucco\ seriously for strong coupling,
then we have for each integer $n$ a supermultiplet of states
of charge $n$ and mass proportional to $|n|$, going to zero mass
as $\lambda\to\infty$. It is very hard to interpret such a spectrum in
terms of local field theory in five dimensions.  But from our previous
experience, we know what to do: interpret these states as Kaluza-Klein
states on ${\bf R}^5\times {\bf S}^1$.

The ${\bf S}^1$ here will have to be a ``new'' circle, not to be
confused with the five-torus ${\bf T}^5$
in the original toroidal compactification
to five dimensions.
(For instance, the $T$-duality group $SO(21,5)$ acts on ${\bf T}^5$
but not on the new circle.)
So altogether, we seem to have eleven dimensions,
${\bf R}^5\times {\bf S}^1\times {\bf T}^5$, and hence we seem
to be in need
of an eleven-dimensional supersymmetric theory.

In section two, eleven-dimensional supergravity made a handy appearance
at this stage, but here we seem to be in a quandary.  There is no obvious
way to introduce an eleventh dimension relevant to the heterotic string.
Have we reached a dead end?

\subsec{The Heterotic String In Six Dimensions}

Luckily, there is a conjectured
relation between the heterotic string and Type II
superstrings \refs{\hull,\vafa} which
has just the right properties to solve our problem
(though not by leading us immediately back to eleven dimensions).
The conjecture is that the heterotic string toroidally
compactified to {\it six} dimensions is equivalent to the Type IIA
superstring compactified to six dimensions on a K3 surface.

The evidence for this conjecture has been that both models have
the same supersymmetry and low energy spectrum
in six dimensions and the same moduli space
of vacua, namely $SO(20,4;{\bf Z})\backslash
 SO(20,4;{\bf R})/(SO(20)\times SO(4))$.
For the toroidally compactified heterotic string, this structure
for the moduli space of vacua is due to Narain \narain;
for Type II, the structure was
determined locally by Seiberg \ref\newsb{N. Seiberg, ``Observations On The
Moduli Space of Superconformal Field Theories,'' Nucl. Phys. {\bf B303} (1988)
286} and globally by Aspinwall and Morrison \ref\aspmor{P. Aspinwall
and D. Morrison, ``String Theory On K3 Surfaces,'' DUK-TH-94-68,
IASSNS-HEP-94/23.}.

In what follows, I will give several new arguments for
this ``string-string duality'' between the heterotic string and
Type IIA superstrings:

(1) When one examines more precisely how the low energy effective actions
match up, one finds that weak coupling of one theory corresponds to
strong coupling of the other theory.  This is a necessary condition
for the duality to make sense, since we certainly know that the
heterotic string for weak coupling is not equivalent to the Type IIA
superstring for weak coupling.

(2) Assuming string-string duality in six dimensions, we will be able
to resolve the puzzle about the strong coupling dynamics of the heterotic
string in five dimensions.  The strongly coupled heterotic string
on ${\bf R}^5$ (times a five-torus whose parameters are kept fixed)
is equivalent to a Type IIB superstring on ${\bf R}^5\times {\bf S}^1$
(times a K3 whose parameters are kept fixed).
The effective six-dimensional
Type IIB theory is weakly coupled at its compactification
scale, so this is an effective solution of the problem of strong coupling
for the heterotic string in five dimensions.

(3) We will also see that -- as anticipated by Duff in a more abstract
discussion \duff\ -- string-string duality in six dimensions implies
$S$-duality of the heterotic string in four dimensions.  Thus, all
evidence for $S$-duality can be interpreted as evidence for string-string
duality, and one gets at least a six-dimensional answer to the question
``what higher dimensional statement leads to $S$-duality in four
dimensions?''

(4) The K3 becomes singular whenever the heterotic string gets
an enhanced symmetry group; the singularities have an $A-D-E$ classification,
just like the enhanced symmetries.

(5) Finally, six-dimensional string-string duality also
leads to an attractive picture for heterotic string dynamics in
seven dimensions.  (Above seven dimensions the analysis would be
more complicated.)

I would like to stress that some of
these arguments test more than a long distance
relation between the heterotic string and strongly coupled Type IIA.
For instance, in working out the five-dimensional dynamics via string-string
duality, we will be led to a Type IIA theory with a {\it small} length
scale, and to get a semi-classical description will require a $T$-duality
transformation, leading to Type IIB.  The validity of the discussion
requires that six-dimensional string-string duality should be an exact
equivalence, like the $SL(2,{\bf Z})$ symmetry
 for Type IIB in ten dimensions and
unlike the relation of Type II to eleven-dimensional supergravity.

\subsec{\it Low Energy Actions}

Let us start by writing a few terms in the low energy effective
action of the heterotic string, toroidally compactified
to six dimensions.  We consider the metric $g$, dilaton $\phi$,
and antisymmetric tensor field $B$, and we let $C$ denote a generic
abelian gauge field arising from the toroidal compactification.
We are only interested in keeping track of how the various terms scale
with $\phi$.  For the heterotic string, the whole classical action
scales as $e^{-2\phi}\sim \lambda^{-2}$, so one has very roughly
\eqn\urggo{I=\int d^6x\sqrt g e^{-2\phi}\left(R+|\nabla\phi|^2+|dB|^2
+|dC|^2\right).}

On the other hand, consider the Type IIA superstring in six dimensions.
The low energy particle content is the same as for the toroidally
compactified heterotic string, at least at a generic point in the moduli
space of the latter where the unbroken gauge group is abelian.
Everything is determined by $N=4$ supersymmetry except the number
of $U(1)$'s in the gauge group and the number of antisymmetric tensor
fields; requiring that these match with the heterotic
string leads one to use Type IIA rather  than Type IIB.
So in particular, the low energy theory derived from Type IIA
has a dilaton $\phi'$, a metric $g'$, an antisymmetric tensor
field $B'$, and gauge fields $C'$.  \foot{We normalize $B'$ and $C'$ to have
standard gauge transformation laws.  Their gauge transformations would
look different if one scaled the fields by powers of $e^\phi$.
This point was discussed in section two.}  Here $\phi'$, $g'$, and $B'$
come from the NS-NS sector, but $C'$ comes from the RR sector, so as
we noted in section two, the kinetic energy of $\phi'$, $g'$, and $B'$ scales
with the dilaton just like that in \urggo, but the kinetic energy of
$C'$ has no coupling to the dilaton.  So we have schematically
\eqn\burggo{I'=\int d^6x\sqrt g' \left(
e^{-2\phi'}\left(R'+|\nabla\phi'|^2+|dB'|^2
\right)+|dC'|^2\right).}

We need the change of variables that turns \urggo\ into \burggo.
In \urggo, the same power of $e^\phi$ multiplies $R$ and $|dC'|^2$.
We can achieve that result in \burggo\ by the change of variables
$g'=g''e^{2\phi'}$.  Then \burggo\ becomes
\eqn\oburggo{I'=\int d^6x\sqrt g'' \left(
e^{2\phi'}(R''+|\nabla\phi'|^2)+e^{-2\phi'}|dB'|^2
+e^{2\phi'}|dC'|^2\right).}
Now the coefficient of the kinetic energy of $B'$ is the opposite
of what we want, but this can be reversed by a duality transformation.
The field equations of $B'$ say that $d*(e^{-2\phi'}dB')=0$, so the
duality transformation is
\eqn\hoborgo{e^{-2\phi'}dB'=*dB''.}
Then \oburggo\ becomes
\eqn\poburggo{I'=\int d^6x\sqrt g'' e^{2\phi'}\left(
R''+|\nabla\phi'|^2+|dB''|^2 +|dC'|^2\right).}
This agrees with \urggo\ if we identify $\phi=-\phi'$.
Putting everything together, the change of variables by which
one can identify the low energy limits of the two theories is
\eqn\yoburggo{\eqalign{ \phi & = -\phi'   \cr
                           g & = e^{2\phi }g' = e^{-2\phi'}g' \cr
                           dB&=e^{-2\phi'}*dB' \cr
                           C & = C'.}}
Unprimed and primed variables are fields of the heterotic string
and Type IIA, respectively.

In particular, the first equation implies that weak coupling of one
theory is equivalent to strong coupling of the other.  This makes
it possible for the two theories to be equivalent without the equivalence
being obvious in perturbation theory.

\subsec{Dynamics In Five Dimensions}

Having such a (conjectured) exact statement in six dimensions,
one can try to deduce the dynamics below six dimensions.
The ability to do this is not automatic because
(just as in field theory) the dimensional reduction might lead to
new dynamical problems at long distances.
But we will see that in this particular case,
the string-string duality in six dimensions does determine what happens
in five and four dimensions.

We first compactify the heterotic string from ten to six dimensions on
a torus (which will be kept fixed and not explicitly mentioned),
and then take the six-dimensional world to be ${\bf R}^5\times
{\bf S}^1_r$, where ${\bf S}^1_r$ will denote a circle of radius $r$.
We want to keep $r$ fixed and take $\lambda=e^\phi$ to infinity.
According to
\yoburggo, the theory in this limit is equivalent to the Type IIA
superstring on ${\bf R}^5\times {\bf S}^1_{r'}$ times a K3 surface (of fixed
moduli),
with string coupling and radius $\lambda'$ and $r'$ given by
\eqn\nybo{\eqalign{\lambda' & = \lambda^{-1} \cr
                   r'       & = \lambda^{-1}r.\cr}}

In particular, the coupling $\lambda'$ goes to zero in the limit
for $\lambda\to \infty$.  However, the radius $r'$ in the dual
theory is also going to zero.   The physical interpretation is much
clearer if one makes a $T$-duality transformation, replacing $r'$ by
\eqn\mybo{r''={1\over r'}={\lambda\over r}.}
The $T$-duality transformation also acts on the string coupling constant.
This can be worked out most easily by noting that the effective
five-dimensional gravitational constant, which is $\lambda^2/r$,
must be invariant under the $T$-duality.
So under $r'\to 1/r'$, the string coupling $\lambda'$
is replaced by
\eqn\bybo{\lambda''= {\lambda'\over r'}}
so that
\eqn\jjn{{r'\over (\lambda')^2}= {r''\over (\lambda'')^2}.}

Combining this with \nybo, we learn that the heterotic string
on ${\bf R}^5\times {\bf S}^1_r$
and string coupling $\lambda$
is equivalent to a Type II superstring
with coupling and radius
\eqn\yiggobo{\eqalign{\lambda'' & = r^{-1} \cr
                         r'' &= {\lambda\over r}.\cr}}
This is actually a Type IIB superstring, since the $T$-duality
transformation turns the Type IIA model that appears in the string-string
duality conjecture in six dimensions into a Type IIB superstring.

\yiggobo\ shows that the string coupling constant of the effective
Type IIB theory remains fixed as $\lambda\to\infty$ with fixed $r$,
so the dual theory is not weakly coupled at all length scales.
However, \yiggobo\ also shows that $r''\to \infty$ in this limit,
and this means that at the length scale of the compactification,
the effective coupling is weak.  (The situation is
similar to the discussion of the strongly coupled
ten-dimensional Type IIA superstring in section two.)
All we need to assume is that the six-dimensional Type II superstring
theory, even with a coupling of order one, is equivalent at long
distances to weakly coupled Type II supergravity.
If that is so, then when compactified on a very large circle,
it can be described at and
above the compactification length by the weakly coupled supergravity,
which describes the dynamics of the light degrees of freedom.

\bigskip
\noindent{\it Moduli Space Of Type IIB Vacua}

The following remarks will aim to give a more fundamental explanation
of \yiggobo\ and a further check on the discussion.

Consider the compactification of Type IIB superstring theory on
${\bf R}^6\times {\rm K3}$.  This gives a chiral $N=4$ supergravity
theory in six dimensions, with five self-dual two-forms (that is,
two-forms with self-dual field strength) and twenty-one anti-self-dual
two-forms (that is, two-forms with anti-self-dual field strength).
The moduli space of vacua of the low energy supergravity theory is
therefore
\ref\romans{L. Romans, ``Self-Duality For Interacting Fields:
Covariant Field Equations For Six-Dimensional Chiral Supergravities,''
Nucl.  Phys. {\bf B276} (1986) 71.}
$G/K$ with $G=SO(21,5)$ and $K$ the maximal subgroup $SO(21)\times SO(5)$.

The coset space $G/K$ has dimension $21\times 5=105$.  The interpretation
of this number is as follows.  There are 80 NS-NS moduli in the conformal
field theory on K3 (that, the moduli space of $(4,4)$ conformal field
theories on K3 is $80$-dimensional).  There are 24 zero modes of RR fields
on K3.  Finally, the expectation value of the dilaton -- the string coupling
constant -- gives one more modulus.  In all, one has $80+24+1=105$
states.  In particular, the string coupling
constant is unified with the others.

It would be in the spirit of $U$-duality to suppose that the Type IIB theory
on ${\bf R}^6\times K3$ has the discrete symmetry group $SO(21,5;{\bf Z})$.
In fact, that follows from the assumption of $SL(2,{\bf Z})$ symmetry
of Type IIB in ten dimensions \hull\ together with the demonstration
in \aspmor\ of a discrete symmetry $SO(20,4;{\bf Z})$ for $(4,4)$
conformal field theories on K3.  For the
$SO(20,4;{\bf Z})$ and $SL(2,{\bf Z})$
do not commute and together generate $SO(21,5;{\bf Z})$.
The moduli space of Type IIB vacua on ${\bf R}^6\times {\rm K3}$ is
hence
\eqn\momo{{\cal N}=SO(21,5;{\bf Z})\backslash SO(21,5;{\bf R})/(SO(21)
\times SO(5)).}

Now consider the Type IIB theory on ${\bf R}^5\times {\bf S}^1\times {\rm
K3}$.  One gets one new modulus from the radius of the ${\bf S}^1$.
No other new moduli appear (the Type IIB theory on ${\bf R}^6\times {\rm K3}$
has no gauge fields so one does not get additional moduli from Wilson lines).
So the moduli space of Type IIB vacua on ${\bf R}^5\times {\bf S}^1
\times {\rm K3}$ is
\eqn\jomo{{\cal M}={\cal N}\times {\bf R}^+,}
where ${\bf R}^+$ (the space of positive real numbers) parametrizes
the radius of the circle.

What about the heterotic string on ${\bf R}^5\times {\bf T}^5$?
The $T$-duality moduli space of the toroidal vacua is
precisely ${\cal N}=
SO(21,5;{\bf Z})\backslash SO(21,5;{\bf R})/(SO(21)
\times SO(5))$.  There is one more modulus, the string coupling constant.
So the moduli space of heterotic string vacua on ${\bf R}^5\times {\bf T}^5$
is once again ${\cal M}={\cal N}\times {\bf R}^+$.  Now the
${\bf R}^+$ parametrizes the string coupling constant.

So the moduli space of toroidal heterotic string vacua on ${\bf R}^5\times
{\bf T}^5$ is the same as the moduli space of Type IIB vacua on
${\bf R}^5\times {\rm K3}$, suggesting that these theories may be equivalent.
The map between them turns the string coupling constant of
the heterotic string into the radius of the circle in the Type IIB
description.
This is the relation that we have seen in \yiggobo\ (so, in particular,
strong coupling of the heterotic string goes to large radius in Type IIB).

To summarize the discussion, we have seen that an attractive conjecture
-- the equivalence of the heterotic string in six dimensions to a certain
Type IIA theory -- implies another attractive conjecture  -- the equivalence
of the heterotic string in five dimensions to a certain Type IIB theory.
The link from one conjecture to the other depended on a $T$-duality
transformation, giving evidence that these phenomena must be understood
in terms of string theory, not just in terms of relations among low
energy field theories.

\bigskip
\noindent{\it Detailed Matching Of States}

Before leaving this subject,
perhaps it would be helpful to be more explicit about how
the heterotic and Type II spectra match up in five dimensions.

Compactification of the six-dimensional heterotic string
theory on ${\bf R}^5\times
{\bf S}^1$ generates in the effective five-dimensional theory
three $U(1)$ gauge fields that were not present in six dimensions.
There is the component $g_{m 6}$ of the metric, the component
$B_{m 6}$ of the antisymmetric tensor field, and the vector
$A_m$ that is dual to the spatial components $B_{mn}$ of
the antisymmetric tensor field.
Each of these couples to charged states: $g_{m 6}$ couples
to elementary string states with momentum around the circle,
$B_{m 6}$ to states that wind around the circle, and $A_m$ to
states that arise as instantons in four spatial dimensions, invariant
under rotations about the compactified  circle.  The mass of these
``instantons'' is $r/\lambda^2$, with the factor of $r$ coming from
integrating over the circle and $1/\lambda^2$ the instanton
action in four dimensions.  The masses of these three classes of states
are hence of order $1/r$, $r$, and $r/\lambda^2$, respectively,
if measured with respect to the string metric.
To compare to Type II, we should remember \yoburggo\ that a Weyl transformation
$g=\lambda^2g'$ is made in going to the sigma model metric of the Type IIA
description.  This multiplies masses by a factor of $\lambda$,
so the masses computed in the heterotic string theory
but measured in the string units of Type IIA are
\eqn\preally{\eqalign{  g_{m 6}: & ~~ {\lambda\over r}\cr
                       B_{m 6}: & ~~ {\lambda r} \cr
                       A_m : & ~~{r\over \lambda}.\cr}}

Likewise, compactification of the six-dimensional Type IIA superstring on
${\bf R}^5\times {\bf S}^1$ gives rise to three vectors
$g'_{m 6}$, $B'_{m6}$, and $A'$.  The first two couple to elementary
string states.  The last presumably couples to some sort of soliton,
perhaps the classical solution that has been called the symmetric
five-brane \callan.  Its mass would be of order
$r'/(\lambda')^2$ in string units for the same reasons as before.
The masses of particles coupling to the three vectors are thus in string
units:
\eqn\oreally{\eqalign{  g'_{m 6}: & ~~ {1\over r'}\cr
                       B'_{m 6}: & ~~ {r'} \cr
                       A'_m : & ~~{r'\over (\lambda')^2}.\cr}}

Now, \preally\ agrees with \oreally\ under the expected
transformation $\lambda=1/\lambda'$, $r=\lambda r'$ provided
that one identifies $g_{m6}$ with $g'_{m6}$;
$B_{m6}$ with $A'_m$; and $A_m$ with $B'_{m6}$.
The interesting point is of course that $B_{m6}$ and $A_m$
switch places.  But this was to be expected from the duality
transformation $dB\sim *dB'$ that enters in comparing the two theories.

So under string-string duality the ``instanton,''
which couples to $A_m$, is turned into the string winding state,
which couples to $B'_{m 6}$, and the string winding state that
couples to $B_{m6}$ is turned into a soliton that couples to $A'_m$.

\subsec{Relation To $S$-Duality}

Now we would like to use six-dimensional string-string duality
to determine the strong coupling dynamics
of the heterotic string in four dimensions.  Once again,
we start with a preliminary toroidal compactification from
ten to six dimensions on a fixed torus that will not be mentioned
further.  Then we take the six-dimensional space to be a product
${\bf R}^4\times {\bf T}^2$, with ${\bf T}^2$ a two-torus.
String-string duality says that this is equivalent to
a six-dimensional Type IIA theory  on ${\bf R}^4\times {\bf T}^2$
(with four extra dimensions in the form of a fixed K3).

One might now hope, as in six and five dimensions, to take the strong
coupling limit and get a useful description of strongly coupled
four-dimensional heterotic string theory in terms of Type II.
This fails for the following reason.  In six dimensions, the duality
related strong coupling of the heterotic string to weak coupling
of Type IIA.  In five dimensions, it related weak coupling of the heterotic
string to coupling of order one of Type IIB (see \yiggobo).  Despite
the coupling of order one, this was a useful description because
the radius of the sixth dimension was large, so (very plausibly) the
effective coupling at the compactification scale is small.  A similar scaling
in four dimensions, however, will show that the strong coupling
limit of the heterotic string in four dimensions is related to
a strongly coupled four-dimensional Type II superstring theory, and now
one has no idea what to expect.

It is remarkable, then, that there is another method
to use six-dimensional string-string duality to determine
the strong coupling behavior of the heterotic string in four dimensions.
This was forecast and explained by Duff \duff\ without reference to
any particular example.  The reasoning
goes as follows.

Recall (such matters are reviewed in
\ref\rab{A. Giveon, M. Porrati, and E.
Rabinovici, ``Target
Space Duality In String Theory,'' hepth/9401139.})
that the $T$-duality group of a two-torus is $SO(2,2)$
which is essentially the same as $SL(2)\times SL(2)$.
Here the two $SL(2)$'s are as follows.  One of them, sometimes
called $SL(2)_U$, acts on the complex structure of the torus.
The other, sometimes called $SL(2)_T$, acts on the combination of the area
$\rho$ of the torus and a scalar $b=B_{56}$ that arises in compactification
of the antisymmetric tensor field $B$.

In addition to $SL(2)_U$ and $SL(2)_T$, the heterotic string in four
dimensions is conjectured to have a symmetry $SL(2)_S$ that acts
on the combination of the four-dimensional string coupling constant
\eqn\huty{\lambda_4=\lambda\rho^{-1/2}}
and a scalar $a$ that is dual to the space-time components $B_{mn}$
($m,n=1\dots 4$).
We know that the heterotic string  has $SL(2)_U$ and $SL(2)_T$ symmetry;
we would like to know if it also has $SL(2)_S$ symmetry.
If so, the strong coupling behavior in four dimensions is determined.

Likewise, the six-dimensional Type IIA theory, compactified
on ${\bf R}^4\times {\bf T}^2$, has $SL(2)_{U'}\times SL(2)_{T'}$
symmetry, and one would like to know if it also has
$SL(2)_{S'}$ symmetry.  Here $SL(2)_{U'}$ acts on the complex structure
of the torus, $SL(2)_{T'}$ acts on the area $\rho'$ and
scalar $b'$ derived from $B'_{56}$, and $SL(2)_{S'}$ would
conjecturally act on the string coupling constant $\lambda_4'$
and the scalar $a'$ that is dual to the ${\bf R}^4$ components
of $B'$.

If string-string duality is correct, then the metrics in the equivalent
heterotic and Type IIA descriptions differ only by a Weyl transformation,
which does not change the complex structure of the torus; hence
$SL(2)_U$ can be identified with $SL(2)_{U'}$.
More interesting is what happens to $S$ and $T$.
Because the duality between the heterotic string and Type IIA
involves $dB\sim * dB'$, it turns $a$ into $b'$ and $a'$ into $b$.
Therefore, it must turn $SL(2)_S$ into $SL(2)_{T'}$ and $SL(2)_{S'}$
into $SL(2)_T$.  Hence the known $SL(2)_T$ invariance of the
heterotic and Type IIA theories implies, if string-string duality
is true, that these theories must also have $SL(2)_S$ invariance!

It is amusing to check other manifestations of the fact that
string-string duality exchanges $SL(2)_S$ and $SL(2)_T$.  For
example, the four-dimensional string coupling $\lambda_4=\lambda\rho^{-1/2}
=\lambda/r$ ($r$ is a radius of the torus)
turns under string-string duality into $1/r'=(\rho')^{-1/2}$.
Likewise $\rho=r^2$ is transformed into $\lambda^2(r')^2
=(r'/\lambda')^2=1/(\lambda'_4)^2$.  So string-string duality
exchanges $\lambda_4$ with $\rho^{-1/2}$,
as it must in order to exchange $SL(2)_S$  and $SL(2)_T$.

\bigskip
\noindent{\it Some Other Models With $S$-Duality}

{}From string-string duality we can not only rederive the familiar
$S$-duality, but attempt to deduce $S$-duality for new
models.  For instance, one could consider in the above a particular
two-torus ${\bf T}^2$ that happens to be invariant under some
$SL(2)_{U}$ transformations, and take the orbifold with respect
to that symmetry group of the six-dimensional heterotic string.
This orbifold can be regarded as a different compactification
of the six-dimensional model, so string-string duality -- if true --
can be applied to it, relating the six-dimensional heterotic
string on this orbifold (and an additional four-torus)
to a Type IIA string on the same orbifold (and an additional K3).

Orbifolding by a subgroup of $SL(2)_{ U}$ does not disturb
$SL(2)_T$, so the basic structure used above still holds; if six-dimensional
string-string duality is valid, then $SL(2)_S$ of the heterotic
string on this particular orbifold follows from $SL(2)_T$ of Type IIA
on the same orbifold, and vice-versa.  This example is of some
interest as -- unlike previously known examples of $S$-duality --
it involves vacua in which supersymmetry is completely broken.
The $S$-duality of this and possible
related examples might have implications in the low energy field theory
limit, perhaps related to phenomena such as those recently uncovered
by Seiberg \gseiberg.

\subsec{Enhanced Gauge Groups}

Perhaps the most striking phenomenon in toroidal compactification of the
heterotic string is that at certain points in moduli space, an enhanced
non-abelian gauge symmetry appears.  The enhanced symmetry
group is always simply-laced and so a product of $A$, $D$, and $E$
groups; in toroidal compactification to six dimensions, one can get
any product of $A$, $D$, and $E$ groups of total rank $\leq 20$.

How can one reproduce this with Type IIA on a K3 surface?
\foot{This question was very briefly raised in section 4.3 of
\ref\ferrara{A. Ceresole, R. D'Auria, S. Ferrara, and A. Van Proeyen,
``Duality Transformations In Supersymmetric Yang-Mills Theories
Coupled To Supergravity,'' hepth-9502072}, and has also been raised
by other physicists.}
It is fairly obvious that one cannot get an enhanced gauge symmetry
unless the K3 becomes singular; only then might the field
theory analysis showing that the RR charges have mass of order
$1/\lambda$ break down.

The only singularities a K3 surface gets are orbifold singularities.
(It is possible for the distance scale of the K3 to go to infinity,
isotropically or not, but that just makes field theory better.)
The orbifold singularities
of a K3 surface have an $A-D-E$ classification.  Any combination of
singularities corresponding to a product of groups with total rank
$\leq 20$ (actually at the classical level the bound is $\leq 19$) can
arise.

Whenever the heterotic string on a four-torus gets an enhanced gauge
group $G$, the corresponding K3 gets an orbifold singularity of type $G$.
This assertion must be a key to the still rather surprising
and mysterious occurrence of extended gauge groups for Type IIA on K3,
so I will attempt to explain it.

The moduli space
\eqn\limbo{{\cal M}=SO(20,4;{\bf Z})\backslash SO(20,4;{\bf R})/(SO(20)
\times SO(4))}
of toroidal compactifications of the heterotic string to six dimensions
-- or K3 compactifications of Type II --
can be thought of as follows.  Begin with a 24 dimensional real vector
space $W$ with a metric of signature $(4,20)$, and containing a
self-dual  even
integral lattice $L$ (necessarily
of the same signature).  Let $V$ be a four-dimensional
subspace of $W$ on which the metric of $W$ is positive definite.
Then ${\cal M}$ is the space of all such $V$'s, up to automorphisms
of $L$.  Each $V$ has a
twenty-dimensional orthocomplement $V^\perp$ on which the metric is
negative definite.

In the heterotic string description, $V$ is the space of charges
carried by right-moving string modes, and $V^\perp$ is the space of charges
carried by left-moving string modes.  Generically, neither $V$ nor $V^\perp$
contains any non-zero points in $L$.  When $V^\perp$ contains such a point $P$,
we get a purely left-moving (antiholomorphic) vertex operator ${\cal O}_P$ of
dimension $d_P=-(P,P)/2$.  (Of course, $(P,P)<0$ as the
metric of $W$ is negative
definite on $V^\perp$.)  $d_P$ is always an integer as the lattice $L$ is even.
The gauge symmetry is extended precisely when $V^\perp$ contains some $P$
of $d_P=1$; the corresponding ${\cal O}_P$ generate the extended gauge
symmetry.

In the K3 description, $W$ is the real cohomology of K3 (including
$H^0$, $H^2$, and $H^4$ together \aspmor).  The lattice $L$ is the lattice
of integral points.  $V$ is the part  of the cohomology generated
by self-dual harmonic forms.  The interpretation is clearest if we restrict
to K3's of large volume, where we can use classical geometry.
Then $H^0$ and $H^4$ split off, and we can take for $W$ the 22 dimensional
space $H^2$, and for $V$ the three-dimensional space of self-dual harmonic
two-forms.

Consider a K3 that is developing an orbifold singularity of type $G$,
with $r$ being the rank of $G$.  In the process,
a configuration of $r$ two-spheres $S_i$ (with an intersection
matrix given by the Dynkin diagram of $G$) collapses to a point.
These two-spheres are holomorphic (in one of the complex
structures on the $K3$), and the corresponding cohomology
classes $[S_i]$ have length squared $-2$.  As they collapse, the
$[S_i]$ become anti-self-dual and thus -- in the limit in which
the orbifold singularity develops -- they lie in $V^\perp$.  (In fact,
as $S_i$ is holomorphic, the condition for $[S_i]$ to be anti-self-dual
is just that it is orthogonal to the Kahler class and so has zero area;
thus the $[S_i]$ lie in $V^\perp$ when and only when the orbifold singularity
appears and the $S_i$ shrink to zero.)  Conversely, the Riemann-Roch
theorem can be used to prove that any point in $V^\perp$
of length squared $-2$ is associated with a collapsed holomorphic
two-sphere.

In sum, precisely when an orbifold singularity of type $G$ appears,
there is in $V^\perp$ an integral lattice of rank $r$, generated by
points of length squared $-2$, namely the $S_i$; the lattice is
the weight lattice of $G$ because of the structure
of the intersection matrix of the $S_i$.  This is the same condition
on $V^\perp$ as the one that leads to extended symmetry group $G$ for the
heterotic string. In the K3 description,
one $U(1)$ factor in the gauge group is associated with each collapsed
two-sphere.  These $U(1)$'s should make up the maximal torus
of the extended gauge group.

Despite the happy occurrence of a singularity -- and so possible
breakdown of field theory -- precisely when an extended gauge group
should appear, the occurrence of extended gauge symmetry in
Type IIA is still rather surprising.  It must apparently mean
that taking the string coupling to zero (which eliminates the RR charges)
does not commute with developing an orbifold singularity (which
conjecturally brings them to zero mass), and that conventional
orbifold computations in string theory correspond to taking the string
coupling to zero first, the opposite of what one might have guessed.

\subsec{Dynamics In Seven Dimensions}

The reader might be struck by a lack of unity between the two parts
of this paper.  In sections two and three, we related Type II superstrings
to eleven-dimensional supergravity.  In the present section, we have
presented evidence for the conjectured relation of Type II superstrings
to heterotic superstrings.  If both are valid, should not eleven-dimensional
supergravity somehow enter in understanding heterotic string dynamics?

I will now propose a situation in which this seems to be true: the strong
coupling limit of the heterotic string in seven dimensions.
I will first propose an answer, and then try to deduce it from
six-dimensional string-string duality.

The proposed answer is that the strong coupling limit of the heterotic
string on ${\bf R}^7\times {\bf T}^3$ gives a theory whose low energy
behavior is governed by
eleven-dimensional supergravity on ${\bf R}^7\times {\rm  K3}$!
The first point in favor of this
is that the moduli spaces coincide.  The moduli space
of vacua of the heterotic string on ${\bf R}^7\times {\bf T}^3$ is
\eqn\imcon{{\cal M}={\cal M}_1\times {\bf R}^+}
with
\eqn\dimcon{{\cal M}_1=
SO(19,3;{\bf Z})\backslash SO(19,3;{\bf R})/SO(19)\times
SO(3).}
Here ${\cal M}_1$ is the usual Narain moduli space, and ${\bf R}^+$
parametrizes the possible values of the string coupling constant.
For eleven-dimensional supergravity compactified on ${\bf R}^7\times {\rm K3}$,
the moduli space of vacua is simply the moduli space of Einstein metrics
on K3.  This does {\it not} coincide with the moduli space of $(4,4)$
conformal field theories on K3, because there is no second rank
antisymmetric tensor field in eleven-dimensional supergravity.
Rather the moduli space of Einstein metrics of volume 1 on K3 is
isomorphic to ${\cal M}_1=
SO(19,3;{\bf Z})\backslash SO(19,3;{\bf R})/SO(19)\times SO(3)$.
\foot{This space
parametrizes three-dimensional subspaces
of positive metric in $H^2({\rm K3},{\bf R})$.
The subspace corresponding to
a given Einstein metric on K3
consists of the part of the cohomology that is
self-dual in that metric.} Allowing the volume to vary gives an extra
factor of ${\bf R}^+$, so that the moduli space of Einstein metrics
on K3 coincides with the moduli space ${\cal M}$ of string vacua.

As usual, the next step is to see how the low energy effective theories
match up.  Relating these two theories only makes sense if
large volume
of eleven-dimensional supergravity (where perturbation theory is good)
corresponds to strong coupling of the heterotic string.
We recall that the bosonic fields
of eleven-dimensional supergravity are a metric $G$ and three-form $A_3$
with action
\eqn\pppilko{I={1\over 2}\int d^{11}x\sqrt G \left(R+|dA_3|^2\right)
          +\int A_3\wedge dA_3\wedge dA_3.}
To reduce on ${\bf R}^7\times {\rm K3}$, we take the eleven-dimensional
line-element to be $ds^2=\tilde g_{mn}dx^mdx^n+e^{2\gamma}h_{\alpha\beta}
dy^\alpha dy^\beta$, with $m,n=1\dots 7$, $\alpha,\beta=1\dots 4$;
here $\tilde g$ is a metric on ${\bf R}^7$, $h$
a fixed metric on K3 of volume 1, and $e^\gamma$ the radius of the K3.
The reduction of $A_3$ on ${\bf R}^7\times {\rm K3}$ gives
on ${\bf R}^7$ a three-form $a_3$, and 22 one-forms that we will generically
call $A$.
The eleven-dimensional Lagrangian becomes very schematically (only keeping
track of the scaling with $e^\gamma$)
\eqn\mpilko{\int d^7x\sqrt{\tilde g}\left(e^{4\gamma}(\tilde R+|d\gamma|^2
+|da_3|^2)+|dA|^2\right).}
To match this to the heterotic string in seven dimensions, we write
$\tilde g = e^{-4\gamma}g$, with $g$ the heterotic string metric in
seven dimensions.  We also make a duality transformation
$e^{6\gamma}da_3=*d B$, with $B$ the two-form of the heterotic string.
Then \mpilko\ turns into
\eqn\umpilko{\int d^7x \sqrt g e^{-6\gamma}\left(R+|d\gamma|^2+
|dB|^2+|dA|^2\right).}
The important point is that the Lagrangian scales with an overall
factor of $e^{-6\gamma}$, similar to the overall factor of $\lambda^{-2}
= e^{-2\phi}$ in the low energy effective action of the heterotic
string.  Thus, to match eleven-dimensional supergravity on ${\bf R}^7
\times {\rm K3}$ with the heterotic string in seven dimensions, one
takes the radius of the K3 to be
\eqn\psnn{e^\gamma=e^{\phi/3}=\lambda^{1/3}.}
In particular, as we hoped,
for $\lambda\to\infty$, the radius of the K3 goes to
infinity, and the eleven-dimensional supergravity theory becomes weakly
coupled at the length scale of the light degrees of freedom.

Now, let us try to show that this picture is a consequence of string-string
duality in six dimensions.  We start with the heterotic string on
${\bf R}^6\times {\bf S}^1\times {\bf T}^3$, where ${\bf S}^1$ is a circle
of radius  $r_1$, and ${\bf T}^3$ is a three-torus that will be held
fixed throughout the discussion.\foot{
Generally, there are also Wilson lines on ${\bf T}^3$ breaking the gauge
group to a product of $U(1)$'s; these will be included with the parameters
of the ${\bf T}^3$ that are kept fixed in the discussion.}
If $\lambda_7$ and $\lambda_6$ denote
the heterotic string coupling constant in seven and six dimensions,
respectively, then
\eqn\hsnnn{{1\over \lambda_6^2}={r_1\over\lambda_7^2}.}
We want to take $r_1$ to infinity, keeping $\lambda_7$ fixed.
That will give a heterotic string in seven dimensions.  Then, after
taking $r_1$ to infinity, we consider the behavior for
large $\lambda_7$, to get
a strongly coupled heterotic string in seven dimensions.

The strategy of the analysis
is of course to first dualize the theory, to a
ten-dimensional Type II theory, and then see what happens to the dual
theory when first $r_1$ and then $\lambda$ are taken large.
Six-dimensional string-string duality says that for fixed $r_1$ and
$\lambda$, the heterotic string on ${\bf R}^6\times {\bf S}^1\times {\bf T}^3$
is equivalent to a Type IIA superstring on ${\bf R}^6\times {\bf K3}$, with
the following change of variables.  The six-dimensional string coupling
constant $\lambda_6'$ of the Type IIA description is
\eqn\yurry{\lambda_6'={1\over \lambda_6} = {r_1^{1/2}\over\lambda_7}.}
The metrics $g$ and $g'$ of the heterotic and Type IIA descriptions are
related by
\eqn\psmms{g=e^{2\phi}g'=\lambda_6^2g'={\lambda_7^2\over r_1}g'.}
In addition, the parameters of the K3 depend on $r_1$ (and the parameters
of the ${\bf T}^3$, which will be held fixed) in a way that we will now
analyze.

There is no unique answer, since we could always apply an $SO(20,4;{\bf Z})$
transformation to the K3.  However, there is a particularly simple answer.
The heterotic string compactified on ${\bf S}^1\times {\bf T}^3$
has 24 abelian gauge fields.  As the radius $r_1$ of the ${\bf S}^1$
goes to infinity, the elementary string states carrying the 24 charges
behave as follows. There is one type of charge (the momentum around
the ${\bf S}^1$) such that the lightest
states carrying only that charge go to zero mass,
with
\eqn\psnopo{M\sim {1\over r_1}.}
There is a second charge, the winding number around ${\bf S}^1$, such
that particles carrying that charge have masses that blow up as $r_1$.
Particles carrying only the other 22 charges have fixed masses in the limit.

Any two ways to reproduce this situation with a K3 will be equivalent
up to a $T$-duality transformation.  There is a particularly
easy way to do this -- take a fixed K3 and scale up the volume $V$,
leaving fixed the ``shape.''  This reproduces the above spectrum
with a relation between $V$ and $r_1$ that we will now determine.

We start with the Type IIA superstring theory in ten dimensions.
The bosonic fields include the metric $g_{10}'$, dilaton $\phi_{10}'$,
gauge field $A$, and three-form $A_3$.  The action is schematically
\eqn\querry{\int d^{10}x\sqrt{g_{10}'}\left(e^{-2\phi_{10}'}R_{10}'
+|dA|^2+|dA_3|^2+\dots\right).}
Upon compactification on ${\bf R}^6\times {\rm K3}$, massless modes
coming from $A$ and $A_3$ are as follows.  $A$ gives rise to a six-dimensional
vector, which we will call $a$.  $A_3$ gives rise to 22
vectors -- we will call them $C_I$ -- and a six-dimensional
three-form, which we will call $a_3$.  If $V$ is the volume of the $K3$,
the effective action in six dimensions scales schematically as
\eqn\uberry{\int d^6x \sqrt {g'}\left({1\over (\lambda_6')^2}R'
+V|da|^2+V|da_3|^2+|dC_I|^2\right).}
Visible in \uberry\ are 23 vectors, namely $a$ and the $C_I$. However,
precisely in six dimensions a three-form is dual to a vector,
by $V da_3=*db$.  So we can replace \uberry\ with
\eqn\bluberry{\int d^6x \sqrt {g'}\left({1\over (\lambda_6')^2}R'
+V|da|^2+{1\over V}|db|^2+|dC_I|^2\right),}
with 24 vectors.  As the canonical kinetic energy of a vector
is
\eqn\luberry{\int d^6x {1\over 4e_{\rm eff}^2}|dA|^2,}
with $e_{\rm eff}$ the effective charge, we see that we have one vector with
effective charge of order
$V^{-1/2}$, one with effective charge of order
$V^{1/2}$, and 22 with effective
charges of order one.

According to our discussion in section two, the mass of a particle
carrying an RR charge is of order $e_{\rm eff}/\lambda_6'$.
So for fixed $\lambda_6'$ and $V\to\infty$, one type of particle goes
to zero mass, one to infinite mass, and 22 remain fixed -- just like the
behavior of the heterotic string as $r_1\to\infty$.
The lighest charge-bearing particle has a mass of order
\eqn\nsnsm{M'={1\over V^{1/2}\lambda_6'}.}
To compare this to the mass \psnopo\ of the lightest particle
in the heterotic string description, we must remember the Weyl transformation
\psmms\ between the two descriptions.  Because of this Weyl transformation,
the relation between the two masses should be $M=\lambda_6^{-1}M'
=\lambda_6'M'$.  So $\lambda_6'$ scales out, and the relation between
the two descriptions involves the transformation
\eqn\bucxx{V=r_1^2.}
The reason that the string coupling constant scales out is that
it does not enter the map between the moduli space of heterotic
string vacua on a four-torus and $(4,4)$ conformal field theories
on K3; the relation \bucxx\ could have been deduced by studying
the description of quantum K3 moduli space in \aspmor\ instead of
using low energy supergravity as we have done.

Since we know from \bucxx\ and \yurry\ how the parameters $V$ and $\lambda_6'$
of the Type  IIA description are related to the heterotic string parameters,
we can identify the ten-dimensional Type IIA string coupling constant
$\lambda_{10}'$, given by
\eqn\piffo{{V\over (\lambda_{10}')^2}={1\over (\lambda_6')^2}.}
We get
\eqn\huffalo{\lambda_{10}'={r_1^{3/2}\over \lambda_7}.}
Thus, for $r_1\to \infty$, the Type IIA theory
is becoming strongly coupled.
At the same time, according to \bucxx\ one has $V\to\infty$, so the Type IIA
theory is becoming decompactified.

In section two, we proposed a candidate for
the strong coupling behavior of Type IIA
on ${\bf R}^{10}$: it is given by eleven-dimensional supergravity
on ${\bf R}^{10}\times {\bf S}^1$.
To be more precise, the relation acted as follows on the massless
modes.  If the line element
of the eleven-dimensional theory is $ds^2=
G^{10}_{ij}dx^idx^j+r_{11}^2(dx^{11})^2
$, $i,j=1\dots 10$, with $G^{10}$ a metric on ${\bf R}^{10}$ and
$r_{11}$ the radius of the circle,
then $r_{11}$ is related to the
ten-dimensional Type IIA string coupling constant by
\eqn\gurry{r_{11}= (\lambda_{10}')^{2/3}={r_1\over\lambda_7^{2/3}}}
and the Type IIA metric $g'$ is related to $G^{10}$ by
\eqn\purry{g'=(\lambda_{10}')^{2/3}G^{10}.}
As this result holds for any fixed metric $g'$ on ${\bf R}^{10}$, it must,
physically, hold on any ten-manifold $M$ as long as the dimensions of
$M$ are scaled up fast enough compared to the growth of the ten-dimensional
string coupling constant.  I will assume that with $\lambda_{10}'$
and $V$ going to infinity as determined above, one is in
the regime in which one can use the formulas \gurry, \purry\ that govern
the strong coupling behavior on ${\bf R}^{10}$.

If this is so, then from \purry\
the volume $V_{11}$ of the K3 using the metric of
the eleven-dimensional supergravity is related to the volume $V$ using
the string metric of the Type IIA description by
\eqn\ruffle{V_{11}=(\lambda_{10}')^{-4/3}
V=\lambda_7^{4/3}r_1^{-2}V=\lambda_7^{4/3}.}

Now we have the information we need to solve our problem.
The heterotic string on ${\bf R}^6\times {\bf S}^1\times {\bf T}^3$,
with radius $r_1$ of the ${\bf S}^1$ and string coupling constant
$\lambda_7$, is related to eleven-dimensional supergravity
on ${\bf R}^6\times {\bf S}^1\times {\rm  K3}$, where the
radius of the ${\bf S}^1$ is given in \gurry\ and the volume of the
K3 in \ruffle.  We are supposed to take the limit $r_1\to\infty$
and then consider the behavior for large $\lambda_7$.
The key point is that $V_{11}$ is independent of $r_1$.
This enables us to take the limit as $r_1\to\infty$; all that happens
is that $r_{11}\to \infty$, so the ${\bf R}^6\times {\bf S}^1\times {\rm K3}$
on which the supergravity theory is formulated becomes ${\bf R}^7\times
{\rm K3}$.  (Thus we see Lorentz invariance between the ``eleventh''
dimension which came from strong coupling and six of the ``original''
dimensions.)  The dependence on the heterotic string coupling $\lambda_7$
is now easy to understand: it is simply that the
volume of the K3 is $V_{11}\sim \lambda_7^{4/3}$.
That is of course the behavior of the volume expected
from \psnn.  So the relation that we have proposed between the heterotic
string in seven dimensions and eleven-dimensional supergravity on
${\bf R}^7\times {\rm K3}$ fits very nicely with the implications
of string-string duality in six dimensions.

\newsec{On Heterotic String Dynamics Above Seven Dimensions}

By now we have learned that the strong coupling dynamics of Type II
superstrings is, apparently, tractable in any dimension and that
the same appears to be true of the heterotic string in dimension
$\leq 7$.  Can we also understand the dynamics of the heterotic
string above seven dimensions?

It might be possible to extend the use of six-dimensional string-string
duality above seven dimensions (just as we extended it above six
dimensions at the end of the last section).  This will require more careful
analysis of the K3's and probably more subtle degenerations
than we have needed so far.

But is there some dual description of the heterotic string above
seven dimensions that would give the dynamics more directly?
For instance, can we find a dual of the heterotic string directly
in ten dimensions?

Once this question is asked, an obvious speculation presents itself,
at least in the case of $SO(32)$.  (For the $E_8\times E_8$ theory
in ten dimensions, I have no proposal to make.)  There is another
ten-dimensional string theory with $SO(32)$ gauge group, namely
the Type I superstring.  Might they in fact be equivalent?
\foot{The $SO(32)$ heterotic string has particles that transform as spinors
of $SO(32)$; these are absent in the elementary string spectrum of
Type I and would have to arise as some sort of solitons if these
two theories are equivalent.}

The low energy effective theories certainly match up; this follows
just from the low energy supersymmetry.  Moreover,
they match up in such a way that strong coupling of one theory
would turn into weak coupling of the other.  This is an essential
point in any possible relation between them, since weak coupling of one
is certainly not equivalent to weak coupling of the other.
In terms of the metric $g$, dilaton $\phi$, two-form $B$, and
gauge field strength $F$, the heterotic string effective action in
ten dimensions scales with the dilaton like
\eqn\hufy{\int d^{10}x\sqrt g e^{-2\phi}\left(R+|\nabla\phi|^2
+F^2+|dB|^2\right).}
If we transform $g=e^\phi g'$ and $\phi=-\phi'$, this scales like
\eqn\bufy{\int d^{10}x\sqrt g'\left( e^{-2\phi'}\left(R'+|\nabla\phi'|^2
\right) +e^{-\phi'}F^2+|dB|^2\right).}
This is the correct scaling behavior for the effective action of the Type I
superstring.  The gauge kinetic energy scales as $e^{-\phi'}$ instead
of $e^{-2\phi'}$ because it comes from the disc instead of the sphere.
The $B$ kinetic energy scales trivially with $\phi'$ in Type I
because $B$ is an RR field.  The fact that $\phi=-\phi'$ means
that strong coupling of one theory is weak coupling of the other, as promised.

Though a necessary condition, this is scarcely strong evidence for
a new string-string duality between the heterotic string and Type I.
However, given that the heterotic and Type II superstrings and
eleven-dimensional supergravity all apparently link up, one would
be reluctant to overlook a possibility for Type I to also enter the story.

Let us try to use this hypothetical new duality to determine the
dynamics of the heterotic string below ten dimensions.
(Below ten dimensions, the $SO(32) $ and $E_8\times E_8$ heterotic
strings are equivalent \ginsparg, so the following discussion applies to both.)
We formulate the heterotic string, with ten-dimensional string
coupling constant $\lambda$,  on ${\bf R}^d\times {\bf T}^{10-d}$
with  ${\bf T}^{10-d}$ a $(10-d)$-torus of radius $r$.  This would
be hypothetically equivalent to a toroidally compactified
Type I theory with coupling
constant $\lambda'=1/\lambda$ and (in view of the Weyl transformation used
to relate the low energy actions) compactification scale $r'=r/\lambda^{1/2}$.
Thus, as $\lambda\to\infty$ for fixed $r$, $\lambda'$ goes to zero,
but $r'$ also goes to zero, making the physical interpretation obscure.
It is more helpful to make a $T$-duality transformation of the Type I theory
to one with radius $r''=1/r'$.
The $T$-duality transformation has a very unusual effect for Type I
superstrings \leigh, mapping them to a system that is actually somewhat
similar to a Type II orbifold; the relation of this unusual orbifold
to the system considered in section four merits further study.
The $T$-duality transformation also changes
the ten-dimensional string coupling constant to a new one $\lambda''$
which obeys
\eqn\jdn{{(r')^{10-d}\over (\lambda')^2}={(r'')^{10-d}\over(\lambda'')^2}}
so that the $d$-dimensional effective Newton constant is invariant.
Thus
\eqn\hdns{\lambda''=\lambda'\left({r''\over r'}\right)^{(10-d)/2}
={\lambda^{(8-d)/2}\over r^{10-d}}.}

So for $d=9$, the strong coupling problem would be completely solved:
as $\lambda\to\infty$ with fixed $r$, $\lambda''\to 0$ (and $r''\to
\infty$, which gives further simplification).  For $d=8$, we have
a story similar to what we have already found in $d=5$ and 7 (and for
Type IIA in $d=10$): though
$\lambda''$ is of order 1, the fact that $r''\to \infty$ means
that the coupling is weak at the compactification scale, so that one
should have a weakly coupled description of the light degrees of freedom.
But below $d=8$, the transformation maps one strong coupling limit
to another.

Of course, once we get down to seven dimensions, we have a conjecture
about the heterotic
string dynamics from the relation to Type II.  Perhaps it is just
as well that the speculative relation of the heterotic string to Type I
does not give a simple answer below eight dimensions.
If there were a dimension in which both approaches could be applied,
then by comparing them we would get a relation between (say) a weakly
coupled Type II string and a weakly coupled Type I string.
Such a relation would very likely
be false, so the fact that the speculative string-string duality
in ten dimensions does not easily determine the strong coupling behavior
below $d=8$ could be taken as a further (weak) hint in its favor.
\bigskip

I would like to thank A. Borel, D. Morrison, R. Plesser and N. Seiberg
for discussions.

\listrefs
\end